\documentclass[preprint,a4paper,superscriptaddress]{revtex4-2}
\setcitestyle{super}

\usepackage[utf8]{inputenc}
\usepackage{amsmath,amsfonts,amssymb}
\usepackage{geometry}
\usepackage{graphicx}
\usepackage{color}
\usepackage{xcolor}
\usepackage{empheq}
\usepackage{hyperref}
\usepackage{bm}
\usepackage{url}
\usepackage{diagbox}
\usepackage{array}
\usepackage{setspace}
\usepackage{multirow}

\newcommand*{\alex}[1]{\textcolor{black}{#1}}

\newcommand{\ug}{\mathbf{u}}

\newcommand{\R}{\mathbf{R}}
\newcommand{\D}{\mathbf{D}}
\newcommand{\C}{\mathbf{C}}

\newcommand{\Ruu}{\mathbf{R}_{\mathbf{uu}}}
\newcommand{\Rrr}{\mathbf{R}_{\bm{\rho \rho}}}

\newcommand{\Rcmp}{R_{\mathcal{M}}}

\newcommand{\thetain}{\boldsymbol{\theta}_{\bm{\textrm{in}}}}

\newcommand{\uin}{\ug_{\textrm{in}}}
\newcommand{\uout}{\ug_{\textrm{out}}}

\newcommand{\rhog}{\bm{\rho}}
\newcommand{\rhogm}{\rhog_{\textrm{m}}}
\newcommand{\zm}{z_{\textrm{m}}}
\newcommand{\rg}{\mathbf{r}}
\newcommand{\rin}{\rg_{\textrm{in}}}
\newcommand{\rout}{\rg_{\textrm{out}}}
\newcommand{\rgp}{\rg_{\textrm{p}}}
\newcommand{\rgm}{\rg_{\textrm{m}}}
\newcommand{\rhoin}{\boldsymbol{\rho}_\textrm{in}}
\newcommand{\rhoout}{\boldsymbol{\rho}_\textrm{out}}

\newcommand{\drho}{{\Delta \bm{\rho}}}
\newcommand{\ie}{\textit{i.e.} }

\begin{abstract}

\textbf{Matrix imaging paves the way towards a next revolution in wave physics. Based on the response matrix recorded between a set of sensors, it enables an optimized compensation of aberration phenomena and multiple scattering events that usually drastically hinder the focusing process in heterogeneous media. Although it gave rise to spectacular results in optical microscopy or seismic imaging, the success of matrix imaging has been so far relatively limited with ultrasonic waves because wave control is generally only performed with a linear array of transducers. In this paper, we extend ultrasound matrix imaging to a 3D geometry. Switching from a 1D to a 2D probe enables a much sharper estimation of the transmission matrix that links each transducer and each medium voxel. Here, we first present an experimental proof of concept on a tissue-mimicking phantom through ex-vivo tissues and then, show the potential of 3D matrix imaging for transcranial applications.}
\end{abstract}

\begin{document}

\title{Three-Dimensional Ultrasound Matrix Imaging}

\author         {Flavien Bureau}
\affiliation    {Institut Langevin, ESPCI Paris, PSL University, CNRS, 75005 Paris, France}
%\email          {flavien.bureau@espci.fr}
\author         {Justine Robin}
\affiliation    {Institut Langevin, ESPCI Paris, PSL University, CNRS, 75005 Paris, France}
\author         {Arthur Le Ber}
\affiliation    {Institut Langevin, ESPCI Paris, PSL University, CNRS, 75005 Paris, France}
\author         {William Lambert}
\affiliation    {Institut Langevin, ESPCI Paris, PSL University, CNRS, 75005 Paris, France}
\affiliation    {Hologic / SuperSonic Imagine, 135 Rue Emilien Gautier,
13290 Aix-en-Provence, France}
\author         {Mathias Fink}
\affiliation    {Institut Langevin, ESPCI Paris, PSL University, CNRS, 75005 Paris, France}
\author         {Alexandre Aubry}
\affiliation    {Institut Langevin, ESPCI Paris, PSL University, CNRS, 75005 Paris, France}
%\email          {alexandre.aubry@espci.fr}

\date{\today}
   \maketitle

   \clearpage 

\noindent {\large \textbf{Introduction}} 

The resolution of a wave imaging system can be defined as the ability to discern small details of an object. In conventional imaging, this resolution cannot overcome the diffraction limit of a half wavelength and may be further limited by the maximum collection angle of the imaging device. However, even with a perfect imaging system, the image quality is affected by the inhomogeneities of the propagation medium. Large-scale spatial variations of the wave velocity introduce aberrations as the wave passes through the medium of interest. Strong concentration of scatterers also induces multiple scattering events that randomize the directions of wave propagation, leading to a strong degradation of the image resolution and contrast. Such problems are encountered in all domains of wave physics, in particular for the inspection of biological tissues, whether it be by ultrasound imaging~\cite{lambert_reflection_2020} or optical microscopy~\cite{ntziachristos_going_2010}, or for the probing of natural resources or deep structure of the Earth's crust with seismic waves~\cite{Yilmaz2001}. 

To mitigate those problems, the concept of adaptive focusing has been adapted from astronomy where it was developed decades ago~\cite{babcock_possibility_1953,roddier_adaptive_1999}. Ultrasound imaging employs array of transducers that allows to control and record the amplitude and phase of broadband wave-fields. Wave-front distortions can be compensated for by adjusting the time-delays added to each emitted and/or detected signal in order to focus ultrasonic waves at a certain position inside the medium~\cite{odonnell_phase-aberration_1988,nock_phase_1989,mallart_adaptive_1994,Ali_2023}. The estimation of those time delays implies an iterative time-consuming focusing process that should be ideally repeated for each point in the field-of-view~\cite{masoy_iteration_2005,montaldo_time_2011}. {Such a complex adaptive focusing scheme cannot be implemented in real time since it is extremely sensitive to motion~\cite{pernot_3-d_2004} whether induced by the operator holding the probe or by the movement of tissues}.%{A distinctive feature of ultrasound imaging is its real-time capability, which makes it very sensitive to movements of the tissue  or the operator holding the probe. In this context, such a complex adaptive focusing scheme cannot be implemented in real time with standard ultrasound imaging systems.} 

Fortunately, this tedious process can now be performed in post-processing~\cite{jaeger_full_2015,chau_locally_2019} thanks to the tremendous progress made in terms of computational power and memory capacity during the last decade. To optimize the focusing process and image formation, a matrix formalism can be fruitful~\cite{varslot_eigenfunction_2004,robert_greens_2008,lambert_distortion_2020,bendjador_svd_2020}. Indeed, once the reflection matrix $\mathbf{R}$ of the impulse responses between each transducer is known, any physical experiment can be achieved numerically, either in a causal or anti-causal way, for any incident beam and as many times as desired. {More specifically, assuming that the medium remains fixed during the acquisition}, a multi-scale analysis of the wave distortions can be performed to build an estimator of the transmission matrix $\mathbf{T}$ between each transducer of the probe and each voxel inside the medium~\cite{lambert_ultrasound_2022}. Once the $\mathbf{T}$-matrix is known, a local compensation of aberrations can be performed for each voxel, thereby providing a confocal image of the medium with a close to ideal resolution and an optimized contrast everywhere. 

Although it gave rise to striking results in optical microscopy~\cite{kang_high-resolution_2017,badon_distortion_2020,yoon_laser_2020,Kwon2023,Najar2023} or seismic imaging~\cite{blondel_matrix_2018,touma_distortion_2021}, the experimental demonstration of matrix imaging has been, so far, less spectacular with ultrasonic waves~\cite{lambert_distortion_2020,bendjador_svd_2020,Sommer2021,lambert_ultrasound_2021}. Indeed, the first proof-of-concept experiments employed a linear array of transducers. Yet, aberrations in the human body are 3D-distributed and a 1D control of the wave-field is not sufficient for a fine compensation of wave-distortions {as already shown by previous works~\cite{ivancevich_phase-aberration_2006,lacefield_time-shift_2001,lindsey_pitch-catch_2013,liu_estimation_1998}}. Moreover, 2D imaging limits the density of independent speckle grains which controls the spatial resolution of the $\mathbf{T}$-matrix estimator~\cite{lambert_ultrasound_2021}. 

In this work, we extend the ultrasound matrix imaging (UMI) framework to 3D using a fully populated matrix array of transducers~\cite{ratsimandresy_3_2002,provost_3d_2014,provost_3-d_2015}. The overall method is first validated by means of a well-controlled experiment combining ex-vivo pork tissues as aberrating layer on top of a tissue-mimicking phantom. 3D UMI is then applied to a head phantom whose skull induces a strong attenuation, aberration and multiple scattering of the ultrasonic wave-field, phenomena that UMI can quantify independently of each other~\cite{lambert_reflection_2020,lambert_ultrasound_2022}. Inspired by the CLASS method developed in optical microscopy~\cite{kang_high-resolution_2017,yoon_laser_2020}, aberrations are here compensated by a novel iterative phase reversal algorithm more efficient for 3D UMI than a singular value decomposition~\cite{robert_greens_2008,lambert_distortion_2020,bendjador_svd_2020}.  In contrast with previous works, the convergence of this algorithm is ensured by investigating the spatial reciprocity between the $\mathbf{T}$-matrices {in transmission and reception}. Throughout the paper, we will compare the gain in terms of resolution and contrast provided by 3D UMI with respect to its 2D counterpart. {In particular, we will demonstrate how 3D UMI can be a powerful tool for optimizing the focusing process inside the brain through the skull.}

\vspace{10 mm}

\noindent {\large \textbf{Results}}

\noindent {\textbf{Beamforming the reflection matrix in a focused basis.}} 
\begin{figure}[tb!]
  \includegraphics[scale=0.7]{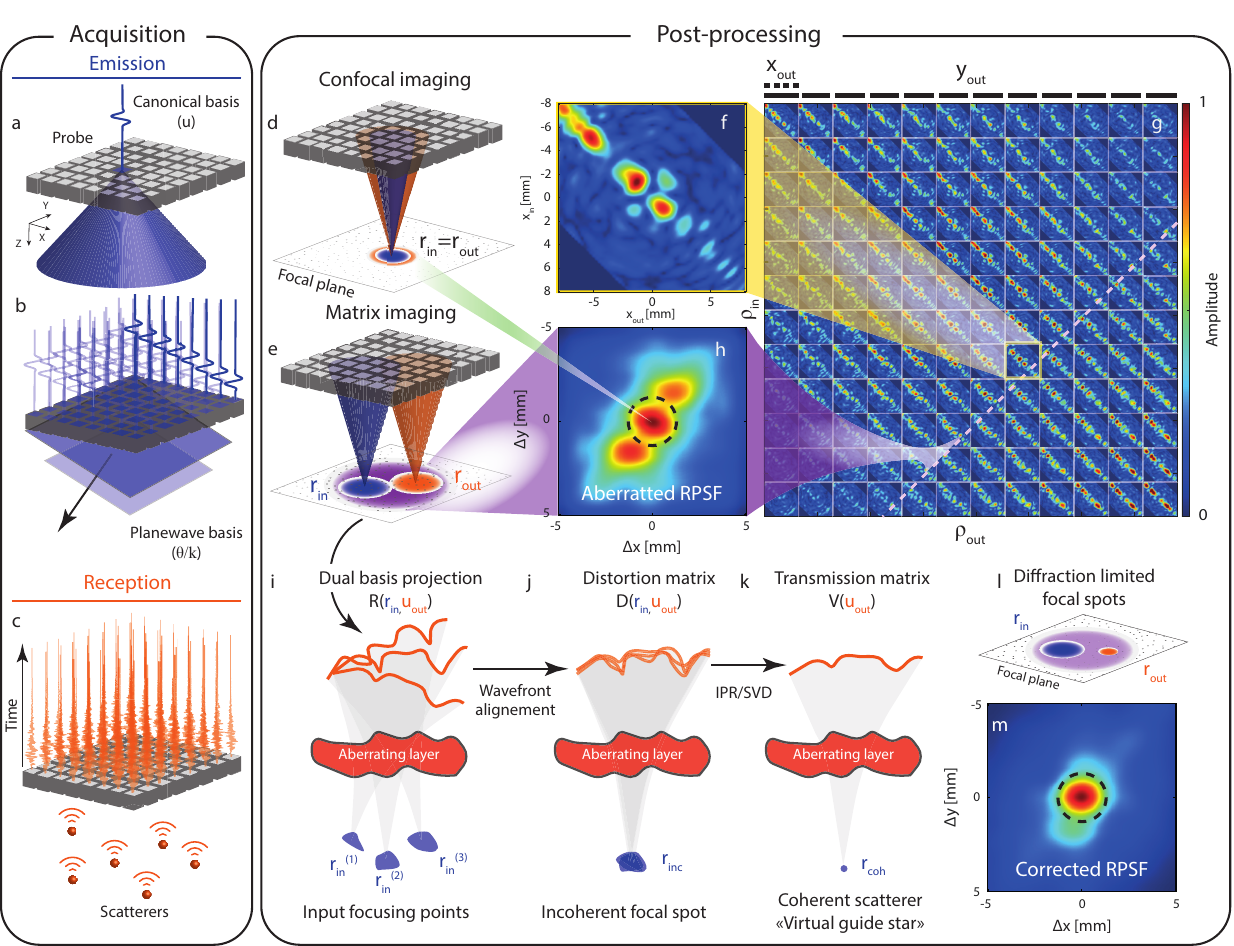}
  \caption{\textbf{3D Ultrasound Matrix Imaging (UMI).} {(\textbf{a},\textbf{b}) The} $\mathbf{R}$-matrix can be acquired in the {transducer} (\textbf{a}) or plane-wave (\textbf{b}) basis in transmit and (\textbf{c}) recording the back-scattered wave-field on each transducer in receive. (\textbf{d}) Confocal imaging consists in a simultaneous focusing of waves at input and output. (\textbf{e}) In UMI, the input ($\rin$) and output ($\rout$) focusing points are decoupled. (\textbf{f}) $x-$cross-section  of the (\textbf{g}) focused $\mathbf{R}-$matrix. (\textbf{h}) UMI enables a quantification of aberrations by extracting a local RPSF (displayed here in amplitude) from each antidiagonal of $\mathbf{R}_{\bm{\rho \rho}}(z)$. (\textbf{i}) UMI then consists in a projection of the focused $\mathbf{R}$-matrix in a correction (here transducer) basis at output. The resulting dual $\mathbf{R}$-matrix connects each focusing point to its reflected wave-front. (\textbf{j}) UMI then consists in realigning those wave-fronts to isolate their distorted component from their geometrical counterpart, thereby forming the $\mathbf{D}$-matrix. (\textbf{k}) An iterative phase reversal algorithm provides an estimator of the $\mathbf{T}-$matrix between the correction basis and the mid-point of input focusing points considered in panel g. (\textbf{l}) The phase conjugate of the $\mathbf{T}-$matrix provides a focusing law that improves the focusing process at output. (\textbf{m}) RPSF amplitude after the output UMI process. The ultrasound data shown in this figure corresponds to the {pork tissue} experiment at depth $z=40$ mm.}
  \label{fig1_3DUMI}
\end{figure}

\begin{figure}[tb!]
  \includegraphics[width=0.9\textwidth]{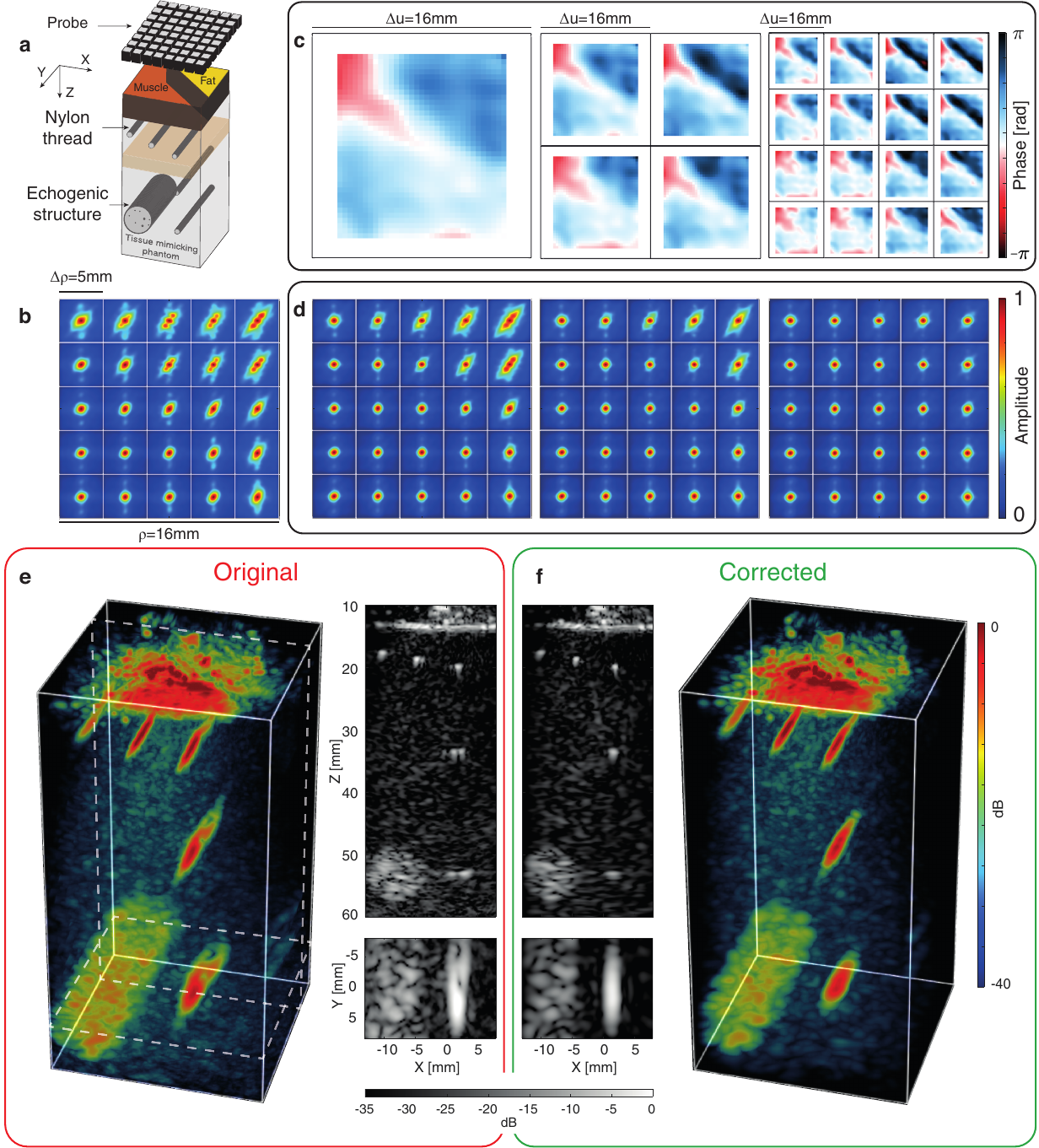}
    \caption{\textbf{Ultrasound matrix imaging of a tissue-mimicking phantom through a {pork tissue}.} (\textbf{a}) Schematic of the experiment. (\textbf{b}) Maps of original RPSFs (in amplitude) at depth $z=29$ mm. (\textbf{c}) Aberration phase laws extracted at the different steps of the UMI process. (\textbf{d}) Corresponding RPSFs after aberration compensation at each step. (\textbf{e},\textbf{f}) 3D confocal and UMI images with one longitudinal and transverse cross-section.}
    \label{fig2_ResultsCDP}
\end{figure}

3D UMI starts with the acquisition of the reflection matrix (see Methods) by means of a 2D array of transducers ($32\times32$ elements, see Fig.~\ref{fig1_3DUMI}a,b). It was performed first on a tissue-mimicking phantom with nylon rods through a {layer of pork tissue of fat and muscle (obtained from a chop rib piece)}, acting as an aberrating layer [Fig.~\ref{fig2_ResultsCDP}a], and then {on a head phantom including brain and skull-mimicking tissue, to reproduce transcranial imaging}  (see below). In the first experiment, the reflection matrix $\mathbf{R}_{\mathbf{u u}}(t)$ is recorded in the {transducer} basis [Fig.~\ref{fig1_3DUMI}a,c], \textit{i.e.} by acquiring the impulse responses, $R(\uin,\uout,t)$, between each transducer ($\ug$) of the probe. In the head phantom experiment, skull attenuation imposes a plane wave insonification sequence [Fig.~\ref{fig1_3DUMI}b] to improve the signal-to-noise ratio. The reflection matrix $\mathbf{R}_{\bm{\theta}\mathbf{ u}}$ then contains the reflected wave-field $R(\bm{\theta}_{\textrm{in}},\uout,t)$ recorded by the transducers $\uout$ [Fig.~\ref{fig1_3DUMI}c] for each incident plane wave of angle $\bm{\theta}_{\textrm{in}}$.

Whatever the illumination sequence, the reflectivity of a medium at a given point $\mathbf{r}$ can be estimated in post-processing by a coherent compound of incident waves delayed to {virtually} focus on this point, and coherently summing the echoes recorded by the probe coming from that same point [Fig.~\ref{fig1_3DUMI}d]. UMI basically consists in decoupling the input ($\rin$) and output ($\rout$) focusing points [Fig.~\ref{fig1_3DUMI}e]. By applying appropriate time delays to the transmission ($\uin / \bm{\theta}_{\textrm{in}}$) and reception ($\uout$) channels ({see Methods}), $\mathbf{R}_{\mathbf{u u}}(t)$ and $\mathbf{R}_{\bm{\theta}\mathbf{u}}(t)$ can be projected at each depth $z$ in a focused basis, thereby forming a broadband focused reflection matrix, $\mathbf{R}_{\bm{\rho \rho}}(z) \equiv [R(\rhoin,\rhoout,z)]$.

Since the focal plane is bi-dimensional, each matrix $\mathbf{R}_{\bm{\rho \rho}}(z)$
has a four-dimension structure: $R(\rhoin,\rhoout,z)= R(\{x_\textrm{in},y_\textrm{in}\},\{x_\textrm{out},y_\textrm{out}\},z)$. $\mathbf{R}_{\bm{\rho \rho}}(z)$ is thus concatenated
in 2D as a set of block matrices to be represented graphically [Fig.~\ref{fig1_3DUMI}g]. In such a
representation, every sub-matrix of $\mathbf{R}$ corresponds to the reflection matrix between lines of virtual transducers located at $y_\textrm{in}$ and $y_\textrm{out}$, whereas every element in the given sub-matrix corresponds to a specific couple $(x_\textrm{in},x_\textrm{out})$ [Fig.~\ref{fig1_3DUMI}e]. Each coefficient $R(x_\textrm{in},y_\textrm{in},x_\textrm{out},y_\textrm{out},z)$ corresponds to the complex amplitude of the echoes
coming from the point $\rout=(x_\textrm{out},y_\textrm{out},z)$ in the focal plane when focusing at point $\rin=(x_\textrm{in},y_\textrm{in},z)$ (or conversely, since $\mathbf{R}_{\bm{\rho \rho}}(z)$ is a symmetric matrix due to spatial reciprocity). 

As already shown with 2D UMI, the diagonal of $\mathbf{R}_{\bm{\rho \rho}}(z)$ directly provides the transverse cross-section of the confocal ultrasound image:
\begin{equation}
    \mathcal{I}(\bm{\rho},z)=|R(\rhoin=\rhoout,z)|^2
\end{equation}
{where $\rhog=\rhoin=\rhoout$ is the transverse coordinate of the confocal point.} The corresponding 3D image is displayed in Fig.~\ref{fig2_ResultsCDP}e for the {pork tissue} experiment. Longitudinal and transverse cross-sections illustrate the effect of the aberrations induced by the {pork} layer by highlighting the distortion exhibited by the image of the deepest nylon rod.

\vspace{5 mm}

\noindent {\textbf{Probing the focusing quality.}} 

We now show how to quantify aberrations in ultrasound speckle (without any guide star) by investigating the antidiagonals of $\Rrr(z)$. In the single scattering regime, the focused $\mathbf{R}-$matrix coefficients can be expressed as follows~\cite{lambert_reflection_2020}:{
\begin{equation}
\footnotesize
    R(\rhoout,\rhoin,z)=\int d\rhog H_{out}(\bm{\rho}-\rhoout,\rhoout,z)\gamma(\bm{\rho},z)H_{in}(\bm{\rho}-\rhoin,\rhoin,z)
\end{equation}}
with $H_\textrm{in/out}$, the input/output point spread function (PSF); and $\gamma$ the medium reflectivity. This last equation shows that each pixel of the ultrasound image (diagonal elements of $\Rrr(z)$) results from a convolution between the sample reflectivity and an imaging PSF, which is itself a product of the input and output PSFs. The off-diagonal points in $\Rrr(z)$ can be exploited for a quantification of the focusing quality at any pixel of the ultrasound image by extracting each antidiagonal. {Such an operation is mathematically equivalent to a change of variable to {express} the focused $\R-$matrix in a common midpoint basis~\cite{lambert_reflection_2020} (see Supplementary Section \alex{2}):
\begin{equation}
    \Rcmp({\drho,\rgm})=R\left(\rhogm-\frac{\drho}{2},\rhogm+\frac{\drho}{2},z\right),
\end{equation}
{where the subscript $\mathcal{M}$ stands for the common midpoint basis.} $\rgm= \left \lbrace\rhogm,z\right \rbrace=\left \lbrace (\rhoin+\rhoout)/2,z\right \rbrace $ is the common midpoint between the input and output focal spots, with the two separated by a distance $\drho=\rhoout-\rhoin$.}

{In the speckle regime (random reflectivity), this quantity probes the local focusing quality as its ensemble average intensity, which we refer to as the \textit{reflection point spread function} (RPSF), scales as an incoherent convolution between the input and output PSFs~\cite{lambert_reflection_2020}:}
\begin{equation}
{{RPSF (\drho,\rgm)}=\left\langle \left\lvert \Rcmp{(\drho,\rgm)}\right\rvert^2\right\rangle \propto 
|H_{\textrm{in}}|^2\overset{\drho}{\circledast} |H_{\textrm{out}}|^2{(\drho,\rgm)} ,}
\end{equation}
{where $\langle \cdots \rangle$ denotes an ensemble average, which, in practice, is performed by a local spatial average (see Methods). }

Figure~\ref{fig1_3DUMI}h displays the mean RPSF associated with the focused $\mathbf{R}-$matrix displayed in Fig.~\ref{fig1_3DUMI}g ({pork tissue} experiment). It clearly shows a distorted RPSF which spreads well beyond the diffraction limit  (black dashed line in Fig.~\ref{fig1_3DUMI}h):
\begin{equation}
\label{resolution}
    \delta \rho_0(z) \sim \frac{\lambda_c}{2\sin \left \lbrace \arctan \left [\Delta u/(2z)\right ] \right \rbrace}
\end{equation}
with $\Delta u$ the lateral extension of the probe. The RSPF also exhibits a strong anisotropy that could not have been grasped by 2D UMI. As we will see in the next section, this kind of aberrations can only be compensated through a 3D control of the wave-field.  

\vspace{5 mm}
\noindent {\textbf{Adaptive focusing by iterative phase reversal.}} 

Aberration compensation in the UMI framework is performed using the distortion matrix concept. Introduced for 2D UMI~\cite{lambert_distortion_2020, lambert_ultrasound_2021}, the distortion matrix can be obtained by: (i) projecting the focused $\mathbf{R}-$matrix either at input or output in a correction basis (here the transducer basis, see Fig. \ref{fig1_3DUMI}i); (ii) extracting wave distortions exhibited by $\mathbf{R}$ when compared to a reference matrix that would have been obtained in an ideal homogeneous medium of wave velocity $c_0$ [Fig. \ref{fig1_3DUMI}j]. The resulting distortion matrix $\mathbf{D}=[D(\mathbf{u},\mathbf{r})]$ contains the aberrations induced when focusing on any point $\mathbf{r}$, expressed in the correction basis.

This matrix exhibits long-range correlations that can be understood in light of isoplanicity. If in a first approximation, the {pork tissue layer} can be considered as a phase screen aberrator, then the input and output PSFs can be considered as spatially invariant: $H_\textrm{in/out}(\bm{\rho}-\bm{\rho}_\textrm{in/out},\mathbf{r}_\textrm{in/out})=H(\bm{\rho}-\bm{\rho}_\textrm{in/out})$. UMI consists in exploiting those correlations to determine the transfer function $T(\mathbf{u})$ of the phase screen. In practice, this is done by considering the correlation matrix $\mathbf{C}=\mathbf{D}\times \mathbf{D}^{\dag}$. The correlation between distorted wave-fields enables a virtual reflector synthesized from the set of output focal spots~\cite{lambert_distortion_2020} [Fig. \ref{fig1_3DUMI}k]. While, in previous works~\cite{lambert_distortion_2020,lambert_ultrasound_2022}, an iterative time-reversal process (or equivalently a singular value decomposition of $\mathbf{D}$) was performed to converge towards the incident wavefront that focuses perfectly through the medium heterogeneities onto this virtual scatterer, here an iterative phase reversal algorithm is employed to build an estimator $\hat{T}(\mathbf{u})$ of the transfer function (see Methods). {Supplementary Figure \alex{3}} demonstrates the superiority of this algorithm compared to SVD for 3D UMI. 

Iterative phase reversal provides an estimation of aberration transmittance {[Fig.~\ref{fig1_3DUMI}k]} whose phase conjugate is used to compensate for wave distortions (see Methods). The resulting mean RPSF is displayed in Fig.~\ref{fig1_3DUMI}m.  Although it shows a clear improvement compared with the initial RPSF, high-order aberrations still subsist. Because of its 3D feature, the {pork tissue layer} cannot be fully reduced to an aberrating phase screen in the transducer basis. 

\noindent {\textbf{Spatial reciprocity as a guide star.}} 

The 3D distribution of the speed-of-sound breaks the spatial invariance of input and output PSFs. Figure~\ref{fig2_ResultsCDP}b illustrates this fact by showing a map of local RPSFs {(see Methods)}.
The RPSF is more strongly distorted below the fat layer of the {pork tissue}  ($c_{\textrm{f}}\approx1480\pm 10$ m/s~\cite{Goss1980}) than below the muscle area ($c_{\textrm{m}}\approx1560 \pm 50$ m/s). A full-field compensation of aberrations similar to adaptive focusing does not allow a fine compensation of aberrations [{Fig.~\ref{fig2_ResultsCDP}d1}]. Access to the transmission matrix $\mathbf{T}=[T(\mathbf{u},\mathbf{r})]$ linking each transducer and each medium voxel is required rather than just a simple aberration transmittance $T(\mathbf{u})$. 

{To that aim, a local correlation matrix $\mathbf{C}(\rgp)$ should be considered around each point $\rgp$ over a sliding box $\mathcal{W}(\rg-\rgp)$ (see Methods), commonly called patches, whose choice of spatial extent {${w}$} is subject to the following dilemma: On the one hand, the spatial window should be as small as possible to grasp the rapid variations of the PSFs across the field of view; on the other hand, these areas should be large enough to encompass a sufficient number of independent realizations of disorder~\cite{robert_greens_2008,lambert_ultrasound_2022}.} The bias made on our $\mathbf{T}$-matrix estimator actually scales as (see Supplementary Section \alex{6}):{
\begin{equation}
\label{eq1}
   | {{\delta {T}(\mathbf{u},\rgp)}}|^2 \sim \frac{1}{\mathcal{C}^2N_{\mathcal{W}}}.
\end{equation}}
{$\mathcal{C}$} is the so-called coherence factor that is a direct indicator of the focusing quality~\cite{mallart_adaptive_1994} but that also depends on the multiple scattering rate and noise background~\cite{lambert_ultrasound_2021}. $N_{\mathcal{W}}$ is the number of {diffraction-limited resolution} cells in each spatial window.

\begin{figure}[tb!]
  \includegraphics[width=0.9\textwidth]{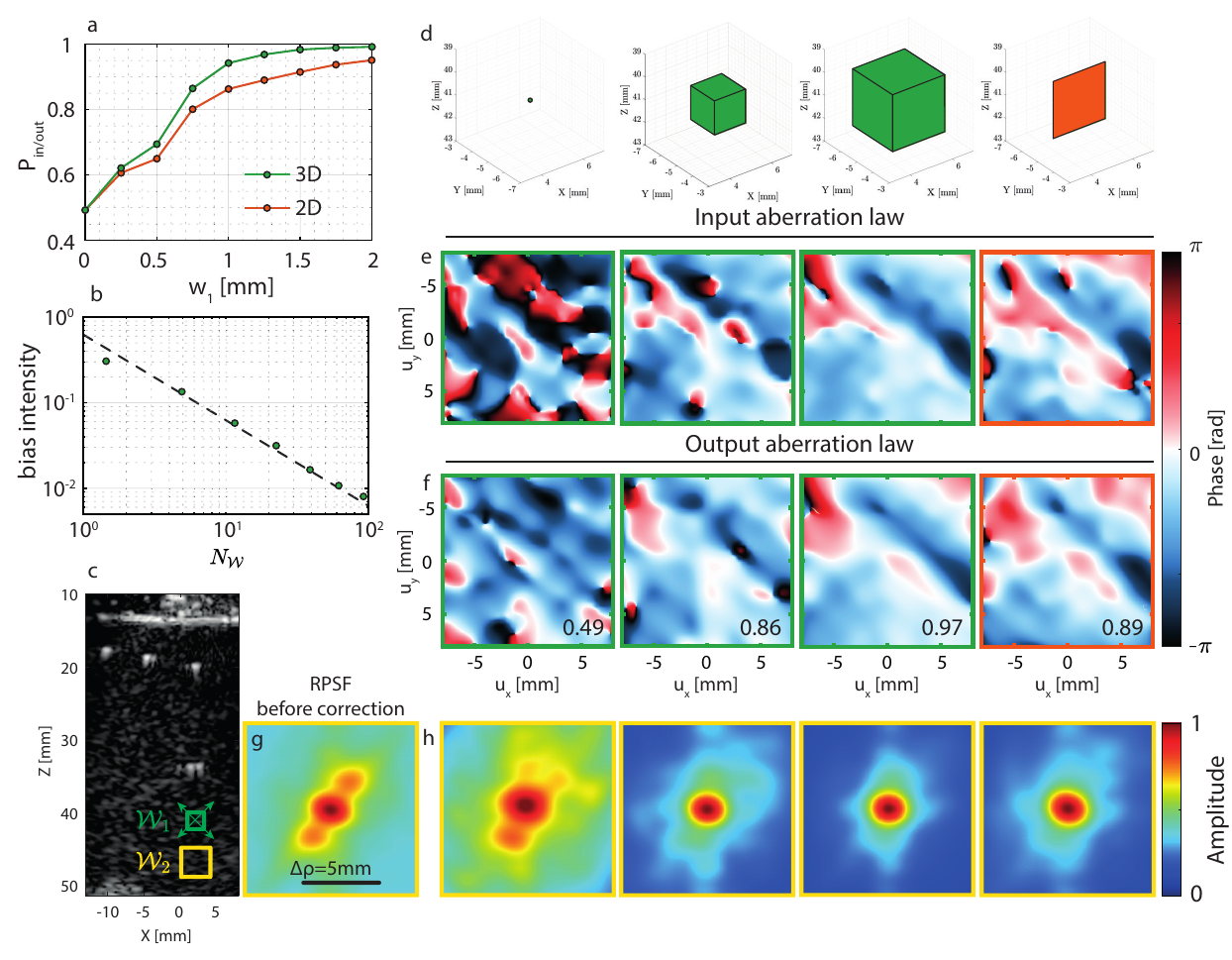}
    \caption{\textbf{Convergence of the UMI process towards the $\mathbf{T}$-matrix.}  (\textbf{a}) Normalized scalar product $P_\textrm{in/out}$ extracted at a point $\rg_1$ (\textbf{c}) as a function of the size ${w}_1$ of the considered spatial window $\mathcal{W}_1$ for 2D (orange) and 3D (green) imaging. (\textbf{b}) Corresponding bias intensity estimator, $|\delta {T}|^2=1-P_{\textrm{in/out}}$, as a function of the number of resolution cells $N_{\mathcal{W}}$ contained in the window $\mathcal{W}_1$. The plot is in log-log scale and the theoretical power law (Eq.~\ref{eq1}) is shown with a dashed black line for comparison. (\textbf{c}) Cross-section of the confocal volume showing the location of $\mathcal{W}_1$ in green and $\mathcal{W}_2$ in yellow. The green box $\mathcal{W}_1$, centered around the point $\rg_1=(5,-5,41)$ mm, denotes the region where the $\mathbf{\hat{T}}-$matrix is extracted, while the yellow box $\mathcal{W}_2$, of fixed size ${w_2}=2$ mm and centered around the point $\rg_2=(5,-5,45)$ mm, is the area where the effect of aberration correction is investigated by means of the RPSF. (\textbf{d}) Spatial windows $\mathcal{W}_1$ considered for the calculation of $\mathbf{C}(\rg_1)$. From left to right: Boxes of dimension $w=0$ mm, $w=0.75$ mm, $w=1.25$ mm, rectangle of dimension $w=1.25$ mm. (\textbf{e},\textbf{f}) Corresponding input $\hat{\mathbf{T}}_\textrm{in}$ and output $\hat{\mathbf{T}}_\textrm{out}$ aberration laws, respectively. The scalar product $P_\textrm{in/out}$ is displayed in each sub-panel of (\textbf{f}). (\textbf{g}) Original RPSF associated with the yellow box $\mathcal{W}_2$ before correction and (\textbf{h}) after correction using the corresponding $\mathbf{\hat{T}}-$matrices displayed in panels (\textbf{e}) and (\textbf{f}). }
    \label{fig4_ConvergenceIPR}
\end{figure}

{{The validity of the $\mathbf{{T}}-$matrix estimator in a region $\mathcal{W}_1$ (Fig.~\ref{fig4_ConvergenceIPR}c) is investigated by examining the corrected RPSF in a neighbour region $\mathcal{W}_2$ (yellow box). $\mathcal{W}_1$ and $\mathcal{W}_2$ are sufficiently close to assume, in a first approximation, that they belong to the same isoplanatic patch.} If the box is too small (left of Fig.~\ref{fig4_ConvergenceIPR}d), our estimator has not converged yet and the correction is not valid, as shown by the degraded quality of the RPSF in $\mathcal{W}_2$ [left panel of Fig.~\ref{fig4_ConvergenceIPR}h] compared {to its initial value}[Fig.~\ref{fig4_ConvergenceIPR}g]. With sufficient spatial averaging [third panel of Fig.~\ref{fig4_ConvergenceIPR}d], a valid aberration law can be extracted, as shown by a corrected RPSF now close to be only diffraction-limited [third panel of Fig.~\ref{fig4_ConvergenceIPR}h].}

{The question that now arises is how we can, in practice, know if the convergence of $\hat{\mathbf{T}}$ is fulfilled without any \textit{a priori} knowledge on ${\mathbf{T}}$. An answer can be found by comparing the estimated input and output aberration phase laws{, $\hat{T}_\textrm{in}(\mathbf{u},\mathbf{r}_{\textrm{p}})$ and $\hat{T}_\textrm{out}(\mathbf{u},\mathbf{r}_{\textrm{p}})$,} at a given point $\mathbf{r}_{\textrm{p}}$ as shown in {Figs.~\ref{fig4_ConvergenceIPR}e and f}. {Spatial reciprocity implies that $\hat{T}_\textrm{in}$ and $\hat{T}_\textrm{out}$ shall be equal when the convergence of the estimator is reached [third panel of Figs.~\ref{fig4_ConvergenceIPR}e and f].} Their normalized scalar product, {$P_\textrm{in/out}=N_u^{-1}\mathbf{\hat{T}}_\textrm{in}\mathbf{\hat{T}}_\textrm{out}^{\dag}$},} {can thus be used to probe the error made on the aberration phase law $| \delta {T} |^2$. Both quantities are actually related as follows (see Supplementary Section \alex{7}): 
\begin{equation}
| \delta {T} |^2 \simeq 1-{P_\textrm{in/out}}.
\end{equation}}
{The normalized scalar product {$P_\textrm{in/out}$} is displayed as a function of {${w}$} and shows the convergence of the IPR process [Fig.~\ref{fig4_ConvergenceIPR}a]. For a sufficiently large box [third panel of Fig.~\ref{fig4_ConvergenceIPR}d], $\hat{\mathbf{T}}$ is supposed to have converged towards ${\mathbf{T}}$ when $\hat{\mathbf{T}}_\textrm{in}$ and $\hat{\mathbf{T}}_\textrm{out}$ are almost equal [third panel of Fig.~\ref{fig4_ConvergenceIPR}e,f], while, for a small box [left panel of Fig.~\ref{fig4_ConvergenceIPR}d], a large discrepancy can be found between them. In the following, {the parameter $P_\textrm{in/out}$} will thus be used as a guide star for {monitoring} the convergence of the UMI process.}
%The error made on the aberration phase law, $\lVert \delta \hat{\mathbf{T}} \rVert^2=1-{N_u^{-1}\mathbf{\hat{T}}_\textrm{in}\mathbf{\hat{T}}_\textrm{out}^{\dag}}$, 
%bis{j'ai enlevé le facteur 2 ici pour être coherent avec ce que j'affiche [Fig 3b], ok avec ca ?}, -> oui j'ai refait le calcul en Supplementary
%can thus be used as an estimator of the bias $\lVert \delta \mathbf{T} \rVert^2$

{{The scaling law of Eq.~\ref{eq1} with respect to $N_{\mathcal{W}}$ is checked {in Fig.~\ref{fig4_ConvergenceIPR}b.}} The inverse scaling of the bias with $N_{\mathcal{W}}$ shows the advantage of 3D UMI over 2D UMI, since $N_{\mathcal{W}}\sim {w}^d$, with $d$ the imaging dimension. This superiority is evident in Fig.~\ref{fig4_ConvergenceIPR}{a}, which shows a faster convergence with 3D boxes (green curve) than with 2D patches (orange curve). For a given precision, 3D UMI thus provides a better spatial resolution for our $\mathbf{T}-$matrix estimator as shown by right panels of Figs.~\ref{fig4_ConvergenceIPR}{f}, where much better agreement between $\mathbf{\hat{T}}_\textrm{in}$ and $\mathbf{\hat{T}}_\textrm{out}$ is observed for a 3D box {[third panel of {Fig.~\ref{fig4_ConvergenceIPR}}d]} than for a 2D patch {[right panel of {Fig.~\ref{fig4_ConvergenceIPR}}d]} of same dimension ${w}$.}

\vspace{5 mm}
\noindent {\textbf{Multi-scale compensation of wave distortions.}} 

The scaling of the bias intensity {$| \delta {T} |^2$} with the coherence factor $\mathcal{C}$ has not been discussed yet. This dependence is however crucial since it indicates that a gradual compensation of aberrations shall be favored rather than a direct partition of the field-of-view into small boxes~\cite{yoon_laser_2020} {(see Supplementary Fig.~\alex{4})}. An optimal UMI process should proceed as follows: first, compensate for input and output wave distortions at a large scale to increase the coherence factor $\mathcal{C}$; then, decrease the spatial window $\mathcal{W}$ and improve the resolution of the $\mathbf{T}-$matrix estimator. The whole process can be iterated, leading to a multi-scale compensation of wave distortions {(see Methods)}. As explained above, the convergence of the process is monitored using spatial reciprocity {($ P_{\textrm{in/out}} >$0.9)}.

The result of 3D UMI is displayed in Fig.~\ref{fig2_ResultsCDP}. It shows the evolution of the $\mathbf{T}-$matrix at each step [Fig.~\ref{fig2_ResultsCDP}c] and the corresponding local RPSFs [Fig.~\ref{fig2_ResultsCDP}d]. %{Note that possible phase wrapping of the estimated aberration laws is not a problem here since aberration correction is performed via their phase conjugation on when used to correct the $\R$-matrix, {since 
%its coefficients are complex IQ signals
%~\cite{kirkhorn_introduction_1999}
%The corresponding time delays, $\tau(\mathbf{u},\mathbf{r}_p)=\mbox{arg} \left \lbrace \hat{T}(\mathbf{u},\mathbf{r}_p) \right \rbrace /(2 \pi f_c) $, are smaller than the pulse duration $\delta t \sim 1$ $\mu$s. 
{In the most aberrated area (\textit{i.e.} under the fat), the phase fluctuations of the aberration law corresponds to a time delay spread of $56$ ns (rms). This value is comparable with past measurements through the human abdominal wall~\cite{hinkelman_measurements_1994}. The pork tissue layer thus induces a level of aberrations typical of standard ultrasound diagnosis.} The comparison with the initial and full-field maps of RPSF highlights the benefit of a local compensation via the $\mathbf{T}-$matrix, with a diffraction-limited resolution reached everywhere. The local aberration phase laws exhibited by {$\hat{\mathbf{T}}$} perfectly match with the distribution of muscle and fat in the {pork tissue layer}. The comparison of the final 3D image [Fig.~\ref{fig2_ResultsCDP}f] and its cross-sections with their initial counterparts [Fig.~\ref{fig2_ResultsCDP}e] show the success of the UMI process, in particular for the deepest nylon rod, which has retrieved its straight shape. The local RPSF on the top right of Fig.\ref{fig2_ResultsCDP} shows a contrast improvement by 4.2 dB and resolution enhancement by a factor 2 ({see Methods and Supplementary Fig.~\alex{5}}).

\vspace{5 mm}
\noindent {\textbf{Overcoming multiple scattering for trans-cranial imaging}} 

\begin{figure}[tb!]
\includegraphics[width=\textwidth]{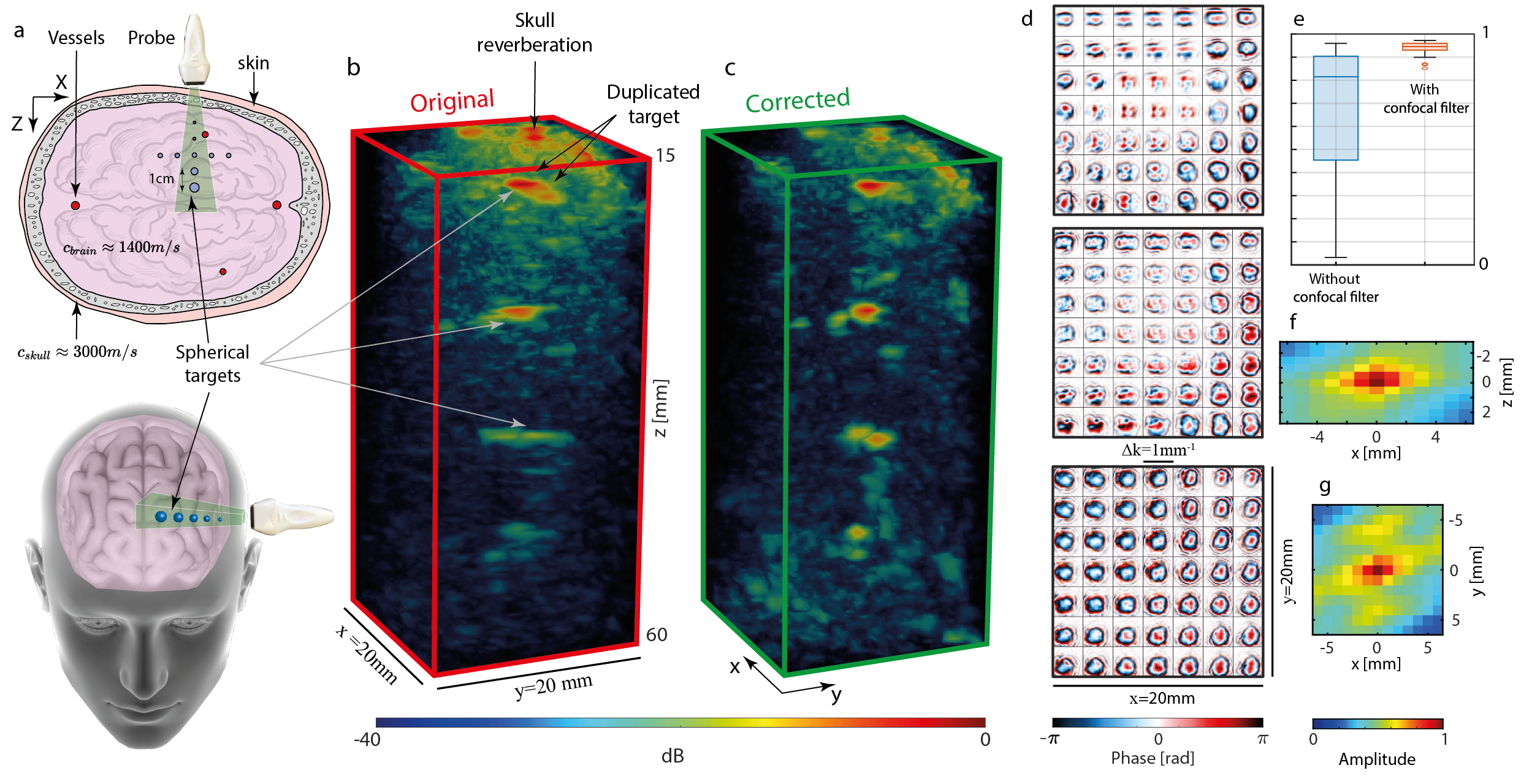}
 \caption{\textbf{Ultrasound Matrix Imaging (UMI) of the head phantom.} (\textbf{a}) Top and oblique views of the experimental configuration\alex{. Image credits: Harryarts and kjpargeter on Freepik}. (\textbf{b,c}) Original and UMI images, respectively. (\textbf{d}) Aberration laws at 3 different depths. From top to bottom: $z=20$ mm, $z=32$ mm, $z=60$ mm. (\textbf{e}) Reciprocity {criterion $P_\textrm{in/out}$} with or without the use of a confocal filter{: Each box chart displays the median, lower and upper quartiles, and the minimum and maximum values.} (\textbf{f},\textbf{g}). Correlation function of the $\hat{\mathbf{T}}$-matrix in the $(x,z)$\alex{-plane (\textbf{f}) and $(x,y)$-plane (\textbf{g})}, respectively. We attribute the sidelobes along the y-axis (\textbf{g}) to the inactive rows separating each block of 256 elements of 
 the matrix array. }
 \label{fig3_HeadPhantomAblaw}
\end{figure}

The same UMI process is now applied to the ultrasound data collected on the head phantom [Fig.~\ref{fig3_HeadPhantomAblaw}a]. The parameters of the multi-scale analysis are provided in the Methods section {(see also Supplementary Fig.~\alex{6})}. The first difference with the {pork tissue} experiment lies in our choice of correction basis. Given the multi-layer configuration in this experiment, the $\mathbf{D}-$matrix is investigated in the plane wave basis~\cite{lambert_distortion_2020}.  

{The second difference is that our spatial reciprocity criterion {$P_\textrm{in/out}$} is very low [{see the} blue box plot in Fig.~\ref{fig3_HeadPhantomAblaw}e]. {This is the manifestation of a bad convergence of our $\mathbf{T}-$matrix estimator.}} The incoherent background exhibited by the original PSFs  [Fig.~\ref{fig5_HeadPhantomPSF}c] drastically affects the coherence factor $\mathcal{C}$~\cite{lambert_ultrasound_2021}, which, in return, gives rise to a strong bias on the $\mathbf{T}-$matrix estimator (Eq.~\ref{eq1}). The incoherent background is due to multiple scattering events in the skull and electronic noise, whose relative weight can be estimated by investigating the spatial reciprocity symmetry of the $\mathbf{R}$-matrix (see Methods). Fig.~\ref{fig5_HeadPhantomPSF}b shows the depth evolution of the single and multiple scattering contributions, as well as electronic noise. {While single scattering dominates at shallow depths ($z<20$ mm), multiple scattering quickly reaches 35\% and remains relatively constant until electronic noise increases, so that the three contributions are almost equal at depths of 75 mm. 

\begin{figure}[tb!]
\includegraphics[width=\textwidth]{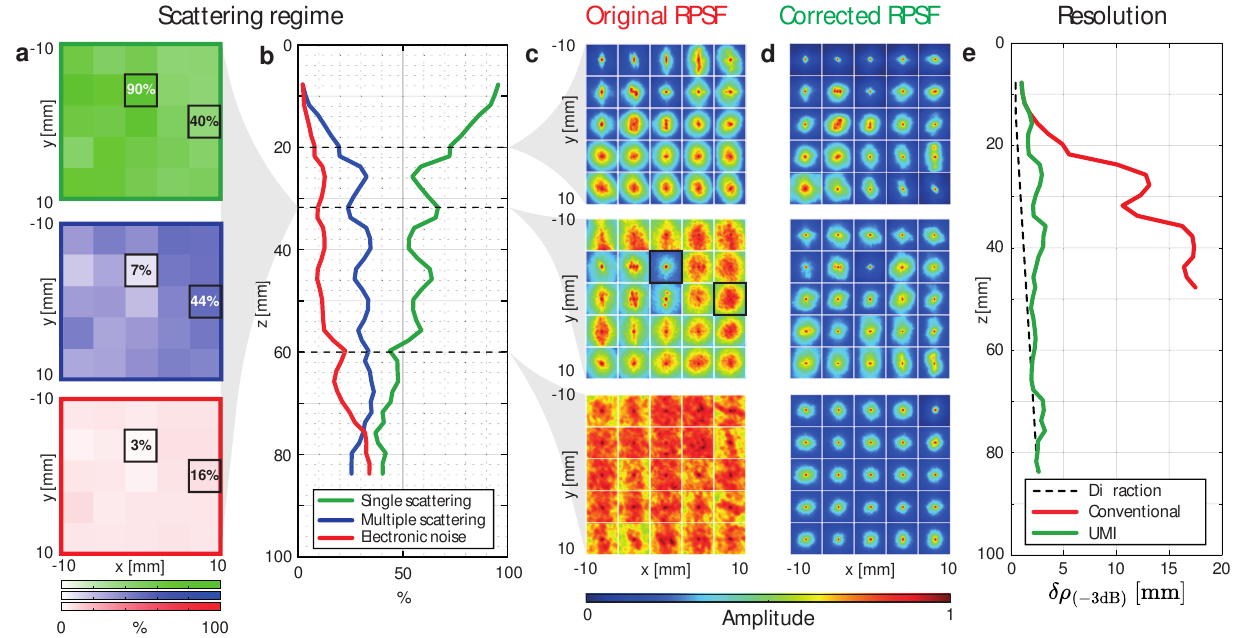}
 \caption{\textbf{Aberrations and multiple scattering quantification.} (\textbf{a}) Single scattering (green), multiple scattering (blue) and noise (red) rate at $z=32$ mm. (\textbf{b}) Single scattering, multiple scattering, and noise rates as a function of depth. (\textbf{c,d}) Maps of local RPSFs (in amplitude) before and after correction, respectively, at three different depths (From left to right: \alex{$z=20$ mm, 32 mm and 60 mm.} Black boxes in panel (\textbf{a}) and (\textbf{c}) corresponds to the same area.
 (\textbf{e}) Resolution {$\delta \rho_{(-3dB)}$} as a function of depth. Initial resolution (red line) and its value after UMI (green line) are compared with the ideal (diffraction-limited) resolution (Eq.~\ref{resolution}).}
 \label{fig5_HeadPhantomPSF}
\end{figure}

Beyond the {depth evolution}, 3D imaging even allows the study of {multiple} scattering in the transverse plane, as shown in Figure \ref{fig5_HeadPhantomPSF}a. Two areas are examined, marked with black boxes, corresponding to the RPSFs shown in [Fig.~\ref{fig5_HeadPhantomPSF}c] ($z=32$ mm). In the center, the RPSFs exhibits a low background due to the presence of a spherical target, resulting in a single scattering rate of 90\%. The second box on the right, however, is characterized by a much higher background, leading to a multiple-to-single scattering ratio slightly larger than one.} This high level of multiple scattering highlights the difficult task of trans-cranial imaging with ultrasonic waves.

In order to overcome these detrimental effects, an adaptive confocal filter can be applied to the focused $\mathbf{R}-$matrix~\cite{lambert_ultrasound_2022}.  
\begin{equation}
R' (\bm{\rho}_\textrm{in},\bm{\rho}_\textrm{out},z) = R (\bm{\rho}_\textrm{in},\bm{\rho}_\textrm{out},z)\exp \left (-\frac{|\bm{\rho}_\textrm{out}-\bm{\rho}_\textrm{in}|^2}{2l_c(z)^2} \right )
\end{equation}
This filter has a Gaussian shape, with a width $l_c(z)$ that scales as $3\delta \rho_0(z)$~\cite{lambert_ultrasound_2022}. The application of a confocal filter drastically improves the correlation between input and output aberration phase laws (see Fig.~\ref{fig3_HeadPhantomAblaw}e {and Supplementary Fig.~\alex{7}}), proof that a satisfying convergence towards the $\mathbf{T}-$matrix is obtained.

Figure~\ref{fig3_HeadPhantomAblaw}d shows the $\mathbf{T}-$matrix obtained at different depths in the brain phantom. {Its spatial correlation function displayed in Figs.~\ref{fig3_HeadPhantomAblaw}f,g provides an estimation of the isoplanatic patch size: 5 mm in the transverse direction (Fig.~\ref{fig3_HeadPhantomAblaw}f) and 2 mm in depth (Fig.~\ref{fig3_HeadPhantomAblaw}g).} This {rapid} variation of the aberration phase law across the field of view confirms {\textit{a posteriori}} the necessity of a local compensation of aberrations induced by the skull. {It also confirms the importance of 3D UMI with a fully sampled 2D array, as previous work recommended that the array pitch should be no more than 50\% of the aberrator correlation length to properly sample the corresponding adapted focusing law~\cite{Lacefield2002}.}

The phase conjugate of the $\mathbf{T}-$matrix at input and output enables a fine compensation of aberrations. A set of {corrected} RPSFs are shown in Fig.~\ref{fig5_HeadPhantomPSF}d. The comparison with their initial values demonstrates the success of 3D UMI: a diffraction-limited resolution is obtained almost everywhere [Fig.~\ref{fig5_HeadPhantomPSF}e)], whether it be in ultrasound speckle  or in the neighborhood of bright targets, at shallow or high depths, which proves the versatility of UMI. 

The performance of 3D UMI is also striking when comparing the three-dimensional image of the head phantom before and  after UMI. [Figs.~\ref{fig3_HeadPhantomAblaw}b and c, respectively]. The different targets were initially strongly distorted by the skull, and are now nicely resolved with UMI. In particular, the first target, located at $z=19$ mm and originally duplicated, has recovered its true shape. In addition, two targets laterally spaced by 10 mm are observed at {42} mm depth, as expected [Fig.~\ref{fig3_HeadPhantomAblaw}a]. {The image of the target observed at {54} mm depth is also drastically improved in terms of contrast and  resolution but is not found at the expected transverse position. One potential explanation is the 
size of this target (2 mm diameter) larger than the resolution cell. The guide star is thus far from being point-like, which can induce an uncertainty on the absolute transverse position of the target in the corrected image.} 

\begin{figure}[tb!]
  \includegraphics[width=\textwidth]{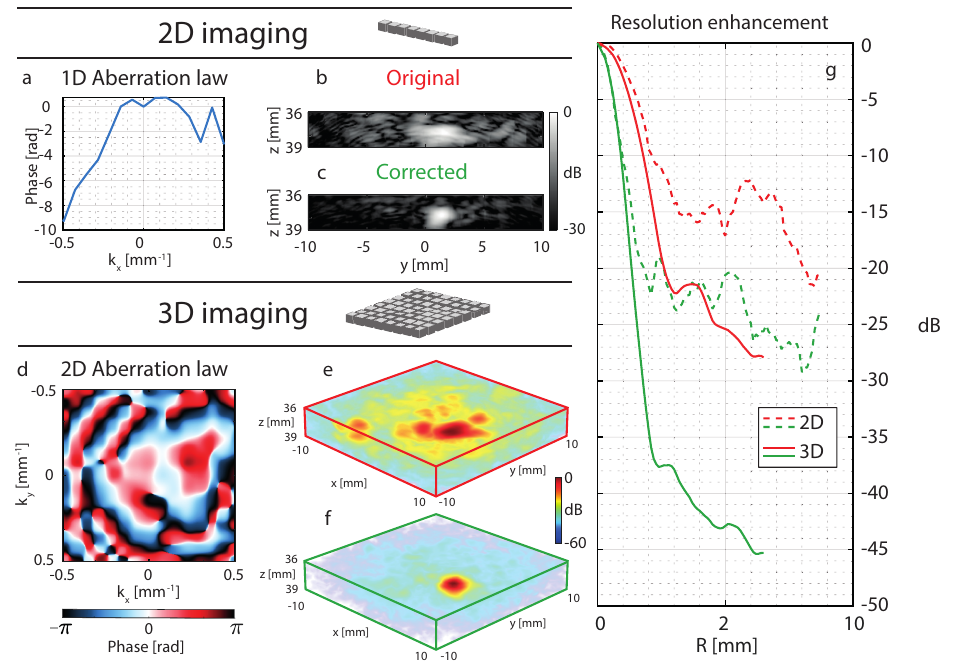}
    \caption{\textbf{2D \textit{versus} 3D matrix imaging in a head phantom.} (\textbf{a}) Aberration law extracted with 2D UMI. (\textbf{b},\textbf{c}) Original and corrected images of the same target with 2D UMI, respectively. (\textbf{d}) Aberration law extracted with 3D UMI for a target located at $z=38$ mm. (\textbf{e},\textbf{f}) Original and corrected images of the same target with 3D UMI, respectively. (\textbf{g})  Imaging PSF before (red) and after (green) 2D (dotted line) and 3D (solid line) UMI.  The depth range considered in each panel corresponds to the echo of the target located at $z=38$ mm.}
    \label{fig6_2Dvs3D}
\end{figure}

Finally, {an isolated target} can be leveraged to highlight the gain in contrast provided by 3D UMI with respect to its 2D counterpart. To that aim, a linear 1D array is emulated from the same raw data by collimating the incident beam in the $y$-direction [Fig. \ref{fig6_2Dvs3D}]. The ultrasound image is displayed before and after UMI in Figs.~\ref{fig6_2Dvs3D}b and c, respectively. The radial average of the corresponding focal spots is displayed in Figs.~\ref{fig6_2Dvs3D}d. Even though 2D UMI enables a diffraction-limited resolution, the contrast gain $G$ is quite moderate ($G_{2D}\sim$ {8dB}) as it scales with the number $N$ of coherence grains exhibited by the 1D aberration phase law [Figs.~\ref{fig6_2Dvs3D}a]: $N_{2D} \sim 6.2$. On the contrary, {as expected}, 3D UMI provides a strong enhancement of the target echo (see the comparison between Figs.~\ref{fig6_2Dvs3D}e,f and g): {$G_{3D}\sim 18$} dB. The 2D aberration phase law actually provides a much larger number of spatial degrees of freedom than its 1D counterpart: $N_{3D}\sim 63$. The gain in contrast is accompanied by a drastic increase of the transverse resolution {($>8\times$ for $z>40$ mm in Fig.~\ref{fig5_HeadPhantomPSF}e)}. Figure \ref{fig6_2Dvs3D} demonstrates the necessity of a 2D ultrasonic probe for trans-cranial imaging. Indeed, the complexity of wave propagation in the skull can only be harnessed with a 3D control of the incident and reflected wave fields.

\newpage

\noindent {\large \textbf{Discussion}}

In this experimental proof-of-concept, we demonstrated the capacity of 3D UMI to correct strong aberrations such as those encountered in trans-cranial imaging. {This work is not only a 3D extension of previous {studies}~\cite{lambert_distortion_2020,lambert_ultrasound_2021}since several crucial elements have been introduced to make UMI more robust.}
%, which is mathematically equivalent to the CLASS algorithm~\cite{kang_high-resolution_2017,yoon_laser_2020} when applied to the whole field-of-view,

{First, the proposed iterative phase reversal algorithm outperforms the SVD for local compensation of aberrations because it can evaluate the aberration law on a larger angular support (see Supplementary Fig.~\alex{3}), {resulting in a sharper compensation of aberrations.}} {Second, the bias of our $\mathbf{T}$-matrix estimator has been expressed analytically (Eq. \ref{eq1}) as a function of the coherence factor that grasps the detrimental effects of the virtual guide star blurring induced by aberrations, multiple scattering and noise. This led us to define a general strategy for UMI with: (\textit{i}) a multi-scale compensation of wave distortions to gradually reduce the blurring of the virtual guide star and tackle high-order aberrations associated with small isoplanatic lengths; (\textit{ii}) the application of an adaptive confocal filter to cope with multiple scattering and noise; (\textit{iii}) a fine monitoring of the convergence of our estimator by means of spatial reciprocity. The latter is a real asset, as it provides an objective criterion to check the physical significance of the extracted aberration laws and optimize the resolution of our $\mathbf{T}-$matrix estimator.}

Although {the results presented in this paper} are striking, they were obtained \textit{in vitro}, and some challenges remain for \textit{in vivo} brain imaging. Until now, UMI has only been applied to a static medium, while biological tissues are usually moving, especially in the case of vascular imaging, where blood flow makes the reflectivity vary quickly over time.  A lot of 3D imaging modes are indeed designed to image blood flow, such as transcranial Doppler imaging~\cite{ivancevich_real-time_2008} or ULM~\cite{bertolo_whole-brain_2021,chavignon_3d_2022}. These methods are strongly sensitive to aberrations~\cite{demene_transcranial_2021,soulioti_super-resolution_2020} and their coupling with matrix imaging would be rewarding to increase the signal-to-noise ratio and improve the image resolution, not only in the vicinity of bright reflectors~\cite{robin_vivo_nodate} but also in ultrasound speckle.

However, due to spatial aliasing, the number of illuminations required for UMI scales with the number of resolution cells covered by the RPSF (see Supplementary Fig.~\alex{8}). Because the aberration level through the skull is important, the illumination basis should thus be fully sampled. It limits 3D transcranial UMI to a compounded frame rate of only a few hertz, which is much too slow for ultrafast imaging~\cite{tanter_ultrafast_2014}. Moreover, a reduced number of illuminations breaks the symmetry of the reflection matrix. It would therefore also affect the accuracy of our monitoring parameter based on spatial reciprocity. 

%{The number of illuminations required for UMI scales with the number of resolution cells covered by the RPSF (see Supplementary Fig.~S7). Because the aberration level through the skull is important, the uncorrected RPSFs are much larger than the diffraction limit.  {This means that the illumination basis should be fully sampled, which limits 3D UMI to a compounded frame rate of only a few hertz, which is much too slow for ultrafast imaging~\cite{tanter_ultrafast_2014}.} In other words, ultrafast acquisition based on a reduced number of illuminations makes the estimation of the $\mathbf{{T}}-$matrix hazardous as spectral aliasing occurs, which may overlap with the underlying RPSFs. In addition, such acquisitions break the symmetry between input and output and thus reduce the accuracy of our spatial reciprocity-based criterion. In this work, such problems did not occur because the complete reflection matrix was acquired with 1225 illuminations.}  
Soft tissues usually exhibit much slower movement, and provide signals several dB higher than blood. Ultrasound imaging of tissues is generally discarded for the brain because of the strong level of aberrations and reverberations. Interestingly, UMI can open a new route towards quantitative brain imaging since a matrix framework can also enable the mapping of physical parameters such as the speed-of-sound~\cite{jaeger_computed_2015,imbault_robust_2017,jakovljevic_local_2018,lambert_reflection_2020}, attenuation and scattering coefficients~\cite{aubry_multiple_2011,Brutt2022}, or fiber anisotropy~\cite{papadacci_ultrasound_2014,rodriguez-molares_specular_2017}. Those various observables can be extremely enlightening for the characterization of cerebral tissues.

{Alternatively, a} solution to directly implement 3D UMI \textit{in vivo} for ultrafast imaging, would be to design an imaging sequence in which the fully sampled $\mathbf{R}-$matrix is acquired prior to the ultrafast acquisition itself, where the illumination basis can be drastically downsampled. {The $\mathbf{\hat{T}}-$matrix} obtained from $\mathbf{R}$ could then be used to correct the ultrafast images in post-processing.

Interestingly, if an ultrafast 3D UMI acquisition is possible (in cases with less aberrations, or at shallow depths), 
the quickly decorrelating speckle observed in blood flow can be an opportunity since it provides a large number of speckle realizations in a given voxel. A high resolution $\mathbf{T}-$matrix could thus be, in principle, extracted without spatial averaging and relying on any isoplanatic assumption~\cite{Zhao1992,Osmanski2012}.  

\alex{So far, one limit of UMI concerns the strong aberration regime in which extreme time delay fluctuations can occur. Indeed, our approach relies on a broadband focused reflection matrix that consists in a coherent time gating of singly-scattered echoes. If time delay fluctuations are larger than the time resolution $\delta t$ of our measurement, the angular components of each echo will not necessarily emerge in the same time gate and aberration compensation will be imperfect.}

{\alex{Beyond strong aberrations,} another issue for transcranial imaging arises from multiple reflections caused by the skull. While {such} %the 
reverberations are not {observed}
%visible 
in the pork tissue experiment, their detrimental effects are much greater in a transcranial experiment because of the large impedance mismatch between the skull and brain tissues. In this work, such artefacts {are} not corrected 
%for 
and they drastically pollute the image at shallow depths ($z<20$ mm).} 

To cope with \alex{those issues}, a polychromatic approach to matrix imaging is required. Indeed, the  aberration compensation scheme proposed in this paper is equivalent to a simple application of time delays on each transmit and receive channel. On the contrary, {a full compensation of reverberation} requires the tailoring of a complex spatio-temporal adaptive (or even inverse) filter. To that aim, 3D UMI provides an adequate framework to exploit, at best, all the spatio-temporal degrees of freedom provided by a high-dimension array of broadband transducers. 

To conclude, 3D UMI is general and can be applied to any insonification sequence (plane wave or virtual source illumination) or array configuration (random or periodic, sparse or dense). Matrix imaging can be also extended to any field of wave physics for which a multi-element technology is available: optical imaging \cite{kang_high-resolution_2017,badon_distortion_2020,yoon_laser_2020}, seismic imaging \cite{touma_distortion_2021,blondel_matrix_2018} and also radar \cite{Berland2020}. All the conclusions raised in that paper can be extended to each of these fields. The matrix formalism is thus a powerful tool for the big data revolution coming in wave imaging.  

\newpage

\noindent {\Large \textbf{Methods}} 

\noindent {\textbf{Description of the {pork tissue} experiment.}}  The first sample under investigation is a tissue-mimicking phantom (speed of sound: $c_0=1540$ m/s) composed of random distribution of unresolved scatterers which generate ultrasonic speckle characteristic of human tissue [Fig.~\ref{fig2_ResultsCDP}a]. The system also contains nylon filaments placed at regular intervals, with a point-like cross-section, and, at a depth of 40 mm, a 10 mm-diameter hyperechoic cylinder, containing a higher density of unresolved scatterers. A 12-mm thick {pork tissue layer} is placed on top of the phantom. It is immersed in water to ensure its acoustical contact with the probe and the phantom. Since the {pork layer} contains a part of muscle tissue ($c_{\textrm{m}} \sim 1560$ m/s) and a part of fat tissue ($c_{\textrm{f}} \sim 1480$ m/s), it acts as an aberrating layer. This experiment mimics the situation of abdominal \textit{in vivo} imaging, in which layers of fat and muscle tissues generate strong aberration and scattering at shallow depths.

The acquisition of the reflection matrix is performed using 
a 2D matrix array of transducers (Vermon) whose characteristics are provided in Tab.~\ref{ProbeInfo}. {The electronic hardware used to drive the probe was developed by Supersonic Imagine (member
of Hologic group) in the context of collaboration agreement with Langevin Institute.}
\begin{table}[!ht]
    \center
\begin{tabular}{|l||l|}
    \hline
  {Number of transducers} & $32 \times 32=1024$ (with 6 dead elements)\\
  \hline
  {Geometry} (y-axis) & 3 inactive rows between each block of 256 elements\\
      \hline
   {Pitch} & $\delta u=0.5$ mm ($\approx \lambda$ at $c=1540$ m/s)\\
   \hline
   {Aperture} & $\Delta \mathbf{u} =\begin{pmatrix}
        \Delta u_x \\  \Delta u_y
    \end{pmatrix} = \begin{pmatrix}
        16 \mbox{ mm} \\  17.5 \mbox{ mm} 
    \end{pmatrix}$ \\
  \hline
  {Central frequency} & $f_c=3$ MHz\\
  \hline
   {Bandwidth} (at $-6$dB) & 80\%$\rightarrow \Delta f=[1.8-4.2]$ MHz\\
    \hline
   {Transducer directivity} & $\theta_{max} =  28^{\circ}$ at  $c=1400$ m/s\\
   \hline
\end{tabular}
\caption{\textbf{Matrix array datasheet.}}
\label{ProbeInfo}
\end{table}

The reflection matrix is acquired by recording the impulse response between each transducer of the probe using IQ modulation with a sampling frequency $f_\textrm{s} = 6 $ MHz. To that aim, each transducer $\uin$ emits successively a sinusoidal burst of three half periods at the central frequency $f_\textrm{c}$.  For each excitation $\uin$, the back-scattered wave-field is recorded by all probe elements $\uout$ over a time length $\Delta t = 139 $ $\mu$s. This set of impulse  responses is stored in the canonical reflection matrix  $\Ruu (t)=[R(\uin , \uout , t)] $.

\vspace{5 mm}

\noindent {\textbf{Description of the head phantom experiment.}}  

In this second experiment, the same probe [Tab.~\ref{ProbeInfo}] is placed slightly above the temporal window of a mimicking head phantom, whose characteristics are described in Tab.~\ref{headphantchar}. To investigate the performance of UMI in terms of resolution and contrast, the manufacturer (True Phantom Solutions) was asked to place small spherical targets made of bone-mimicking material inside the brain. They are arranged crosswise, evenly spaced in the 3 directions with a distance of 1 cm between two consecutive targets, and their diameter increases with depth: $0.2$, $0.5$, $1$, $2$, $3$ mm [Fig.~\ref{fig3_HeadPhantomAblaw}a]. Skull thickness is of $\sim 6$ mm on average at the position where the probe is placed  and the first spherical target is located at $z\approx20$ mm depth, while the center of the cross is at $z\approx40$ mm depth. The transverse size of the head is $\sim 14$ cm.

\begin{table}[!ht]
   \center
   \small
\begin{tabular}{|c|c|c|c|c|c|c|}
    \hline
    & {Speed-of-sound} & {Density}  & {Attenuation}\\
     & [m/s] &  [g/cm$^3$] & @2.25 MHz [dB/cm]\\
  \hline
  Cortical bone & $3000 \pm 30$ & 2.31 & $6.4 \pm 0.3$ \\
  \hline
  Trabecular bone & $2800 \pm 50$ & 2.03 & $ 21 \pm 2$ \\
  \hline
  Brain tissue & $1400 \pm 10$ & 0.99  & $1.0 \pm 0.2$ \\
  \hline
  Skin tissue & $1400 \pm 10$ & 1.01  & $1.7 \pm 0.2$ \\
   \hline
\end{tabular}
\caption{\textbf{Head phantom characteristics.}}
\label{headphantchar}
\end{table}

To improve the signal-to-noise ratio, the $\mathbf{R}$-matrix is here acquired using a set of plane waves~\cite{montaldo_coherent_2009}. For each plane wave of angles of incidence $\thetain=(\theta_x,\theta_y)$, the time-dependent reflected wave field  $R(\thetain,\uout,t)$ is recorded by each transducer $\uout$. This set of wave-fields forms a reflection matrix acquired in the plane wave basis, $\mathbf{R}_{\bm{\theta}\mathbf{u}} (t)=\left [ R(\bm{\theta}_{\textrm{in}},\uout,t) \right ]$. Since the transducer and plane wave bases are related by a simple Fourier transform at the central frequency, the array pitch $\delta u$ and probe size $\Delta u$ dictate the angular pitch $\delta \theta$ and maximum angle $\theta_{max}$ necessary to acquire a full reflection matrix in the plane wave basis such that: $\theta_\textrm{max}=\arcsin [ \lambda_c/(2\delta u)]\approx 28^\circ$; $\delta \theta =\arcsin \left [ \lambda_c/(2\Delta u_y) \right ] \approx 0.8^{\circ}$, with $\lambda_c=c_0/f_c$ the central wavelength and $c_0=1400$ m/s the speed-of-sound in the brain phantom. A set of 1225 plane waves are thus generated by applying appropriate time delays {$\Delta \tau (\bm{\theta}_\textrm{in},\uin)$} to each transducer $\uin=(u_x,u_y)$ of the probe:
\begin{equation}
{\Delta \tau (\bm{\theta}_\textrm{in},\uin)} = [ u_x \sin \theta_x+u_y \sin \theta_y ]  /{c_0}. 
\end{equation}

\vspace{5 mm}
\noindent {\textbf{Focused beamforming of the reflection matrix.}} The focused $\mathbf{R}-$matrix, $\bm{R}_{\bm{\rho\rho}}(z)=[R(\rhoin , \rhoout,z)] $, is built in the time domain via a conventional delay-and-sum beamforming scheme that consists in applying appropriate time-delays in order to focus at different points at input {$\rin=(\rhoin,z)=(\{x_{\textrm{in}},y_{\textrm{in}}\},z)$ and output $\rout=(\rhoout,z)=(\{x_{\textrm{out}},y_{\textrm{out}}\},z)$:
\begin{multline}
\label{beam}
R(\rhoin,\rhoout,z)=\sum_{\mathbf{i}_\textrm{in}}\sum_{\uout}A(\{\mathbf{i}_\textrm{in},\rin\},\{\uout,\rout\})R\left(\mathbf{i}_\textrm{in},\uout,\tau(\mathbf{i}_\textrm{in},\rin)+\tau(\uout,\rout)\right)
\end{multline}
where $\mathbf{i}=\mathbf{u}$ or $\bm{\theta}$ accounts for the illumination basis. $A$ is an apodization factor that limit the extent of the synthetic aperture at emission and reception.} This synthetic aperture is dictated by the transducers' directivity $\theta_\textrm{max} \sim 28^{\circ}$~\cite{perrot_so_2021}. 

In the transducer basis, the {time-of-flights, $\tau(\ug,\rg)$}, writes:
\begin{equation}
   { \tau(\ug,\rg)}=\frac{|\mathbf{u}-\mathbf{r}|}{c_0}=\frac{\sqrt{(x-u_x)^2+(y-u_y)^2+z^2}}{c_0}.
\end{equation}
In the plane wave basis, {$\tau(\bm{\theta},\rg)$} is given by 
\begin{equation}
    {\tau(\bm{\theta},\rg)}=\left [ x \sin \theta_x+ y \sin \theta_y +z \sqrt{1 - \sin^2 \theta_x -\sin^2 \theta_y } \right]  /{c_0}. 
\end{equation}

\vspace{5 mm}
\noindent {\textbf{Local average of the reflection point spread function.}} To probe the local RPSF, {the field-of-view is divided into spatial regions $\mathcal{W}({\rgm}-\rgp)$, defined by their center $\rgp$ and their extent $\mathbf{w}=(w_\rho,w_z)$, where $w_\rho$ and $w_z$ denote the lateral and axial extent, respectively.} A local average of the back-scattered intensity can then be performed in each region:{
\begin{equation}
    RPSF(\Delta \bm{\rho},\rgp)={\left \langle \left\lvert R_{\mathcal{M}}{(\drho,\rgm)}\right\rvert^2 \mathcal{W}(\rg_\mathrm{m}- \rgp) \right \rangle_{\rg_\mathrm{m}}}
\end{equation}
where {the symbol} $\langle \cdots \rangle$ denotes here a spatial average {over the variable in the subscript}. $\mathcal{W}(\rgm-\rgp) = 1$ for $|\rhogm - \rhog_{\textrm{p}}|<w_\rho {/2}$ and $|\zm-z_{\textrm{p}}|<w_z{/2}$, and zero otherwise. The dimensions of $\mathcal{W}$ used for [Fig.~\ref{fig2_ResultsCDP}b,d] are: $\mathbf{w}=(w_\rho,w_z)=(3.2,3)$ mm. The dimensions of $\mathcal{W}$ to obtain [Figs.~\ref{fig5_HeadPhantomPSF}c,d] are: $\mathbf{w}=(w_\rho,w_z)=(4,5.5)$ mm.}\\

\vspace{5 mm}
\noindent {\textbf{Distortion Matrix in 3D UMI}}. The first step consists in projecting the focused $\mathbf{R}-$matrix $\Rrr(z)$ [Fig.~\ref{fig1_3DUMI}e] onto a dual basis {$\mathbf{c}$} at output [Fig. \ref{fig1_3DUMI}i]: 
\begin{equation}
    \R_{\bm{\rho}\mathbf{c}} {(z)}=  \Rrr(z) \times {\mathbf{G}_{\bm{ \rho}\mathbf{c}}}(z)
\end{equation}
where {the symbol} $\times$ {stands for the}  matrix product. {${\mathbf{G}_{\bm{\rho}\mathbf{c}}(z)}$ is the propagation} matrix predicted by the {homogeneous} propagation model {between the focused basis ($\bm{\rho}$) and the correction basis ($\mathbf{c}$) at each depth $z$}. {$\mathbf{c}$ can be either the plane wave, the transducer, or any other correction basis suitable for a particular experiment \cite{fink_aberration_1997,mertz_field_2015,Kwon2023}.}

In the transducer basis {($\mathbf{c} = \mathbf{u}$)}, {the coefficients of ${\mathbf{G}_{\bm{\rho}\mathbf{u}}(z) }$} correspond to the $z-$derivative of the Green's function~\cite{lambert_ultrasound_2022}:
\begin{equation}
{G(\bm{\rho},\mathbf{u},z)=\frac{ze^{ik_c \sqrt{|\mathbf{u}-\bm{\rho}|^2+z^2}}}{4\pi (|\mathbf{u}-\bm{\rho}|^2+z^2)}}
\end{equation}
where $k_c$ is the wavenumber at the central frequency. In the Fourier basis ($\mathbf{c} = \mathbf{k}$), {{$\mathbf{G}_{\bm{\rho}\mathbf{k}}$}} simply corresponds to the Fourier transform operator~\cite{lambert_distortion_2020}:
\begin{equation}
 {G(\bm{\rho},\mathbf{k})} =\textrm{exp}\left(j\mathbf{k}.\bm{\rho})=\textrm{exp}(j\left(k_xx+k_yy\right)\right).
\end{equation}

At each depth $z$, the reflected wave-fronts contained in $\R_{\bm{\rho}\mathbf{c}}$ are then decomposed into the sum of a geometric component ${\mathbf{G}_{\bm{\rho}\mathbf{c}}}$, that would be ideally obtained in absence of aberrations, and a distorted component that corresponds to the gap between the measured wave-fronts and their ideal counterparts [Fig.~\ref{fig1_3DUMI}j] \cite{lambert_distortion_2020,lambert_ultrasound_2022}:
\begin{equation}
   {\mathbf{D}_{\bm{\rho}\mathbf{c}} (z)  =  \mathbf{G}_{\bm{\rho}\mathbf{c}}^*(z)} \circ  \mathbf{R}_{\bm{\rho}\mathbf{c}}(z)
\end{equation}
where the symbol $\circ$ stands for a Hadamard product. {$\mathbf{D}_{\mathbf{r}\mathbf{c}}=\mathbf{D}_{\bm{\rho}\mathbf{c}} (z)=[D(\{\rhoin,z\},\mathbf{c}_\textrm{out})]$ is the so-called distortion matrix, here expressed at the output. }{Note that the same operations can be performed by exchanging input and output to obtain the input distortion matrix $\mathbf{D}_{\mathbf{c}\mathbf{r}}=[D(\mathbf{c}_\textrm{in},\rg_{\textrm{{out}}})]=[D(\mathbf{c}_\textrm{{in}},\{\rhog_{\textrm{{out}}},z\})]$.}

\vspace{5 mm}
\noindent {\textbf{Local correlation analysis of the $\mathbf{D}-$matrix}}. The next step is to exploit local correlations in {${\mathbf{D}}_{\rg\mathbf{c}}$} to extract the $\mathbf{T}$-matrix. To that aim, a set of {output} correlation matrices $\mathbf{C}_{\textrm{out}}(\rgp)$ shall be considered between distorted wave-fronts in the vicinity of each point $\mathbf{r}_\textrm{p}$ in the field-of-view:
\begin{equation}
\label{corr_in}
    C ({\mathbf{c}_{\textrm{out}}} ,{\mathbf{c}'_{\textrm{out}}},\rgp)= \left \langle {D}(\rg_\textrm{in},{\mathbf{c}_\textrm{out}}){D}^*(\rg_\textrm{in},{\mathbf{c}'_\textrm{out}}) \mathcal{W}(\rg_\textrm{in}- \rgp)\right \rangle _{\rg_\textrm{in}}
\end{equation}
An equivalent operation can be performed in input in order to extract a local correlation matrix ${\mathbf{C}_{\textrm{in}}}(\mathbf{r}_\textrm{p})$ from the input distortion matrix $\mathbf{D}_{{\mathbf{c}}\mathbf{r}}$.

\vspace{5 mm}
\noindent {\textbf{Iterative phase reversal algorithm}}. The iterative phase reversal algorithm is a computational process that provides an estimator of the transmission matrix, 
\begin{equation}
{{\mathbf{T}}_\textrm{out}(z)=\mathbf{G}^{{{\top}}}_{{\bm{\rho} \mathbf{c}}}(z) \times {\mathbf{H}}_\textrm{out}}{(z)},
\end{equation}
where the superscript {$\top$} stands for matrix transpose. %and $\mathbf{G}_{{\mathbf{r} \bm{o}}}=[G(\lbrace \bm{\rho_\textrm{out},z}\rbrace,\bm{o}_\textrm{out})]$. 
{${\mathbf{T}}_\textrm{out}$}{=[T($\mathbf{c}_\textrm{out},\mathbf{r}_\mathrm{p}$)]} links each point {$\mathbf{c}_\textrm{out}$} in the dual basis and each voxel $\mathbf{r}_\mathrm{p}$ of the medium to be imaged [Fig.~\ref{fig1_3DUMI}k]. Mathematically, the algorithm is based on the following recursive relation:
\begin{equation}
\hat{\mathbf{T}}_\textrm{out}^{(n)} (\mathbf{r}_{\textrm{p}}) = \exp\left[i \, \mathrm{arg}\left\{  \mathbf{C}_{{\textrm{out}}}(\mathbf{r}_{\textrm{p}}) \times {\hat{\mathbf{T}}}_\textrm{out}^{(n-1)} (\mathbf{r}_{\textrm{p}}) \right\}\right]
\end{equation}
where $\hat{\mathbf{T}}_\textrm{out}^{(n)}$ is the estimator of ${\mathbf{T}}_\textrm{out}$ at the $n^\textrm{th}$ iteration of the phase reversal process.  $ \hat{\mathbf{T}}_\textrm{out}^{(0)}$ is an arbitrary wave-front that initiates the iterative phase reversal process (typically a flat phase law) and $\hat{\mathbf{T}}_\textrm{out}= \lim_{n\to\infty} \hat{\mathbf{T}}_\textrm{out}^{(n)}$ is the result of this iterative phase reversal process. 

This iterative phase reversal algorithm, repeated for each point $\mathbf{r}_\mathrm{p}$, yields an estimator $\hat{\mathbf{T}}_\textrm{out}$ of the $\mathbf{T}$-matrix. Its digital phase conjugation enables a local compensation of aberrations [Fig.~\ref{fig1_3DUMI}l]. The focused $\mathbf{R}-$matrix can be updated as follows:
\begin{equation}
\label{dopc}
     \mathbf{R}^{{\textrm{(corr)}}}_{\bm{\rho} \bm{\rho}}(z)=     \left [   \mathbf{D}_{\bm{\rho}\mathbf{c}} (z)\circ {\hat{\mathbf{T}}^\dag_\textrm{out}}(z)\right ]\times  {\mathbf{G}_{\bm{\rho}\mathbf{c}}^{\dag}}(z)
\end{equation}
where the symbol $\dag$ stands for transpose conjugate and $\circ$ for the Hadamard product. The same process is then applied to the input correlation matrix $\mathbf{C}_{{\textrm{in}}}$ for the estimation of the input transmission matrix, {$\mathbf{T}_\textrm{in}{(z)}=\mathbf{G}_{{{\bm{\rho}\mathbf{c}}} }^{{\top}}{(z)}\times {\mathbf{H}}_\textrm{in}{(z)}$.}% bis{Remplacer par ??? $\rightarrow$ $\mathbf{T}_\textrm{in}{(z)}= \mathbf{G}_{{\mathbf{c}\bm{\rho}} }{(z)} \times {\mathbf{H}}_\textrm{in}$}

\vspace{5 mm}
\noindent {\textbf{Multi-scale analysis of wave distortions.}} To ensure the convergence of the IPR algorithm, several iterations of the aberration correction process are performed while reducing the size of the patches $\mathcal{W}$ with an overlap of 50\% between them. Three correction steps are performed in the {pork tissue} experiment, whereas six are performed in the head phantom experiment [as described in Table \ref{patchsize}]. At each step, the correction is performed both at input and output and reciprocity between input and output aberration laws is checked. The correction process is stopped if the normalized scalar product {$P_\textrm{in/out}$} does not reach $0.9$. 
\begin{table}[h!]
\center
\begin{tabular}{|c||c|c|c||c|c|c|c|c|c|}
\hline
 & \multicolumn{3}{c||}{{{Pork tissue}}} & \multicolumn{6}{c|}{{Head phantom}}\\
\hline
\hline
{Correction step} & $1^{\circ}$ & $2^{\circ}$ & $3^{\circ}$ & $1^{\circ}$ & $2^{\circ}$ & $3^{\circ}$ & $4^{\circ}$ & $5^{\circ}$ & $6^{\circ}$\\
\hline
Number of transverse patches & $1 \times 1$ & $2 \times 2$ & $4 \times 4$ & $1 \times 1$ & $2 \times 2$ & $3 \times 3$ & $4 \times 4$ & $5 \times 5$ & $6 \times 6$\\
\hline
$w_{\bm{\rho}} =(w_x,w_y)$ [mm] & 16 & 12 & 8 & 20 & 15 & 13.3 & 10 & 8 & 6.6\\
\hline
$w_z$ [mm] & 3 & 3 & 3 & 5.5 & 5.5 & 5.5 & 5.5 & 5.5 & 5.5 \\
\hline
\end{tabular}
\caption{\textbf{Parameters of UMI in both experiments.}}
\label{patchsize}
\end{table}

\vspace{5 mm}
\noindent {\textbf{Synthesise a 1D linear array.}}
To estimate the benefits of 3D imaging compared to 2D UMI, a simulation of a 1D array is performed on experimental ultrasound data acquired with our 2D matrix array. To that aim, cylindrical time delays are applied at input and output:
\begin{equation}
    \tau'(\theta^{(s)},s,z)=\frac{ s \sin \theta^{(s)} +z \cos \theta^{(s)} }{c_0}
\end{equation}
\begin{equation}
   \tau'({u}^{(s)},s,z)=\frac{\sqrt{(s-u^{(s)})^2+z^2}}{c_0}.
\end{equation}
with $s=x$ or $y$, depending on our focus plane choice.

The focused $\mathbf{R}-$matrix is still built in the time domain but using this time the following delay-and-sum beamforming:
\begin{align}
\footnotesize
 R^{(2D)}(y_\textrm{in},y_\textrm{out},z)=\sum_{\bm{\theta}_\textrm{in}}\sum_{\uout} &  {R}\left(\thetain,\uout,\overbrace{\tau'(\theta_\textrm{in}^{(y)},y_\textrm{in},z)+ \tau'(u_\textrm{out}^{(y)},y_\textrm{out},z)}^{{\textrm{2D beamforming along (\textit{y,z})-plane}}} \right. \nonumber \\
& \left . +\underbrace{\tau'(\theta_\textrm{in}^{(x)},x_\textrm{f},z_\textrm{f})+\tau'(u_\textrm{out}^{(x)},x_\textrm{f},z_\textrm{f}) -2z_\textrm{f}/c_0}_{{\textrm{Cylindrical law to focus at ($x_\textrm{f},z_\textrm{f}$)}}}\right ).
\label{equ_loicylindrique}
\end{align}
The images displayed in {Fig.~\ref{fig6_2Dvs3D}b,c} are obtained by synthesizing input and output beams collimated in the $(y,z)-$plane by focusing on a line located at ($x_\textrm{f}=0$ mm, $z_\textrm{f}=37.25$ mm), thereby mimicking the beamforming process by a conventional linear array of transducers.

\vspace{5 mm}
\noindent {\textbf{Estimation of contrast and resolution.}} 
Contrast and resolution are evaluated by means of the RPSF. Equivalent to the full width at half maximum commonly used in 2D UMI, the transverse resolution $\delta \rho $ is assessed in 3D based on the area $\mathcal{A}_{(-3dB)}$ at half maximum of the RPSF amplitude:
\begin{equation}
\delta \rho_{(-3dB)}=\sqrt{\mathcal{A}_{(-3dB)}/\pi}
\end{equation}
The contrast, {$\mathcal{F}$}, is computed locally by decomposing the {normalised RPSF} as the sum of three components~\cite{lambert_ultrasound_2021}: 
\begin{equation}
\label{decomposition}
    {\overline{RPSF}(\mathbf{r}_{\textrm{p}},\Delta \bm{\rho})=\frac{RPSF(\mathbf{r}_{\textrm{p}},\Delta \bm{\rho})}{RPSF(\mathbf{r}_{\textrm{p}},\Delta \bm{\rho}=\mathbf{0})}=\alpha_S(\mathbf{r}_{\textrm{p}})+\alpha_M(\mathbf{r}_{\textrm{p}})+ \alpha_N (\mathbf{r}_{\textrm{p}}).}
\end{equation}
$\alpha_S$ is the single scattering {rate} that corresponds to {the} confocal peak. $\alpha_M$ is a multiple scattering {rate} that gives rise to a diffuse halo; $\alpha_N$ corresponds to the electronic noise {rate} which results in a flat plateau. A local contrast can then be deduced from the ratio between $\alpha_S$ and the incoherent background {$\alpha_B=\alpha_M+\alpha_N$},
\begin{equation}
{ \mathcal{F}}(\mathbf{r}_{\textrm{p}})={\frac{ \alpha_S(\mathbf{r}_{\textrm{p}})}{\alpha_B(\mathbf{r}_{\textrm{p}})}=\frac{1-\alpha_B(\mathbf{r}_{\textrm{p}})}{\alpha_B(\mathbf{r}_{\textrm{p}})}}
 \label{contrastdef}
\end{equation}

\vspace{5 mm}
\noindent {\textbf{Single and multiple scattering rates.}} The single scattering, multiple scattering and noise rates can be directly computed from the decomposition of the RPSF (Eq.~\ref{decomposition}). However, at large depths, multiple scattering and noise are difficult to discriminate since they both give rise to a flat plateau in the RPSF. In that case, the spatial reciprocity symmetry can be invoked to differentiate their contribution. The multiple scattering component actually gives rise to a symmetric $\mathbf{R}$-matrix
while electronic noise is associated with a fully random matrix. The relative part of the two components can thus be estimated by computing the degree of {anti-symmetry} {$\beta$} in the $\mathbf{R}-$matrix. {To that aim, the $\mathbf{R}$-matrix is first projected onto its anti-symmetric subspace at each depth :
\begin{equation}
    \mathbf{R}^{\textrm{(A)}}_{\bm{\rho \rho}}(z)=\frac{\mathbf{R}_{\bm{\rho\rho}}(z)-\mathbf{R}_{\bm{\rho\rho}}^\top(z)}{2}
    \label {antisymdef}
\end{equation}
where the superscript $\top$ stands for matrix transpose. In a common midpoint representation, (Eq. \ref{antisymdef}) re-writes: 
\begin{equation}
    \Rcmp^{\textrm{(A)}}(\rgm,\drho)=\frac{\Rcmp(\rgm,\drho)-\Rcmp(\rgm,-\drho)}{2}.
\end{equation}
A local degree of anti-symmetry $\beta$ is then computed as follows:
\begin{equation}
    \beta(\mathbf{r_{\textrm{p}}})=\frac{\left\langle\left\lvert \Rcmp^{\textrm{(A)}}(\rgm,\drho)\right\rvert^2\mathcal{W}(\mathbf{r}_{\textrm{m}}-\mathbf{r}_{\textrm{p}})\mathcal{D}(\Delta \bm{\rho})\right\rangle_{[\rgm,\drho]}}{\left\langle\left\lvert \Rcmp(\rgm,\drho)\right\rvert^2\mathcal{W}(\mathbf{r}_{\textrm{m}}-\mathbf{r}_{\textrm{p}})\mathcal{D}(\Delta \bm{\rho})\right\rangle_{[\rgm,\drho]}}
    \label{symmetryratedef}
\end{equation}
{where $\mathcal{D}(\Delta \bm{\rho})$ is a de-scanned window function that eliminates the confocal peak such that the computation of $\beta$ is only made by considering the incoherent background. Typically, we chose $\mathcal{D}(\Delta \bm{\rho})=1$ for $\Delta \bm{\rho}>6\delta \rho_0(z)$, and zero otherwise.} Assuming equi-partition of the electronic noise between its symmetric and anti-symmetric subspace, the multiple scattering rate $\alpha_M$ and noise ratio $\alpha_N$ can then be deduced {(see Supplementary Section \alex{11})}:
\begin{align}
  &  \alpha_M(\mathbf{r}_{\textrm{p}})=\left(1-2\beta(\mathbf{r}_{\textrm{p}})\right) \alpha_{B}(\mathbf{r}_{\textrm{p}})\\
   & \alpha_N(\mathbf{r}_{\textrm{p}})=2\beta(\mathbf{r}_{\textrm{p}})  \alpha_{B}(\mathbf{r}_{\textrm{p}})
\end{align}
In the head phantom experiment [Fig.~\ref{fig5_HeadPhantomPSF}b], these rates are estimated at each depth by averaging over a window of size $\mathbf{w}=(w_\rho,w_z)=(20,5.5)$ mm.}

\vspace{5 mm}
\noindent {\textbf{Computational insights.}} While the UMI process is close to real-time for 2D imaging (\ie for linear, curve or phased array probes), 3D UMI {(using a fully populated matrix array of transducers)} is still far from it (see Tab.~\ref{computational}) as it involves the processing of much more ultrasound data. Even if computing a confocal 3D image only requires a few {minutes}, building the focused $\mathbf{R}-$matrix from the raw data takes a few hours (on GPU with CUDA language) while one step of aberration correction only lasts for a few {minutes}. All the post-processing was realized with Matlab (R2021a) on a working station with 2 processors @2.20GHz, 128Go of RAM, and a GPU with 48 Go of dedicated memory.\\

\begin{table}[htb]
\begin{tabular}{|c|c|c|c|c|c|}
  \hline
  \multicolumn{2}{|c|}{} &
  \multicolumn{2}{c|}{{2D} imaging} &
  \multicolumn{2}{c|}{{3D} imaging} \\
  \hline
  \multicolumn{2}{|c|}{Number of channels [Input $\times$ Output]} & \multicolumn{2}{c|}{$32 \times 32 \approx 10^3$} & \multicolumn{2}{c|}{$1024 \times 1024\approx 10^6$}\\
    \hline
        \multicolumn{2}{|c|}{Field-of-view $(\Delta x,\Delta y,\Delta z)$} & \multicolumn{2}{c|}{$(20,0,80)$ mm} & \multicolumn{2}{c|}{$(20,20,80)$ mm} \\ 
            \hline
  \multicolumn{2}{c|}{} & {Data} & {Time} & {Data} & {Time}\\ 
  \hline
  \multicolumn{2}{|c|}{Reflection matrix acquisition:  $\Ruu(t)$} & \multicolumn{1}{c|}{$6$ Mo} & \multicolumn{1}{c|}{$8$ ms} & \multicolumn{1}{c|}{$6$ Go} & \multicolumn{1}{c|}{$260$ ms} \\ 
  \hline
  \multicolumn{2}{|c|}{Confocal image $\mathcal{I}(\mathbf{r})$} & $53$ ko & $5.1$ ms & $2.2$ Mo & $1.3$ min \\ 
  \hline

  \multirow{2}{*}{{Matrix Imaging}}
    & Focused $\mathbf{R}-$matrix: $\mathbf{R}_{\bm{\rho \rho}}(z)$ & $2.2$ Mo & $15$ ms & $3.6$ Go & $2.3$ h \\ 
    & Estimation of ${\mathbf{T}}$ \& correction & \ & $0.15$ s & \ & $4.5$ min\\
   \hline
\end{tabular}
\caption{\textbf{Computational insights.} Here, we compare the typical amount of data and computational time at each post-processing step of UMI. The comparison between 2D and 3D imaging is made using a single line of transducers \textit{versus} all the transducers of our matrix array. In both cases, the pixel/voxel resolution is fixed at $0.5$ mm, which corresponds approximately to one wavelength. The maximum distance between the input and output focusing points is set to $10$ mm. The estimation of {$\mathbf{{T}}$} is here investigated without a multi-scale analysis on a single iteration at input and output.}
\label{computational}
\end{table}

\clearpage

\noindent\textbf{Data availability.} \alex{The ultrasound data generated in this study is available at Zenodo~\cite{Bureau_2023} ({\href{https://zenodo.org/record/8159177}{https://zenodo.org/record/8159177}}).}\\

\noindent\textbf{Code availability.}
Codes used to post-process the ultrasound data within this paper are available from the corresponding author upon request.\\

\noindent\textbf{Acknowledgments.}
{The authors wish to thank L. Marsac for providing initial ultrasound acquisition sequences.} \alex{The authors are grateful for the funding provided by the European Research Council (ERC) under the European Union's Horizon 2020 research and innovation program (grant agreement 819261, REMINISCENCE project, AA).} \\

\noindent\textbf{Author Contributions.}
A.A. and M.F. initiated the project. A.A. supervised the project. F.B. and A.L.B. coded the ultrasound acquisition sequences. F.B. and J.R. performed the experiments. F.B., A.L.B., and W.L. developed the post-processing tools. F.B., J.R. and A.A. analyzed the experimental results. A.A. performed the theoretical study. F.B. prepared the figures. F.B., J.R. and A.A. prepared the manuscript. F.B., J.R., {A.L.B.}, W.L., M.F., and A.A. discussed the results and contributed to finalizing the manuscript. \\

\noindent\textbf{Competing interests.}
A.A., M.F., and W.L. are inventors of a patent related to this work held by CNRS (no. US11346819B2, published May 2022). W.L. had his PhD funded by the SuperSonic Imagine company and is now an employee of this company. All authors declare that they have no other competing interests.

\vspace{\baselineskip}

\clearpage 

\clearpage

\renewcommand{\thetable}{S\arabic{table}}
\renewcommand{\thefigure}{S\arabic{figure}}
\renewcommand{\theequation}{S\arabic{equation}}
\renewcommand{\thesection}{S\arabic{section}}

\setcounter{equation}{0}
\setcounter{figure}{0}
\setcounter{section}{0}

\begin{center}
\Large{\bf{Supplementary Information}}
\end{center}
\normalsize
%\section*{Subhead}
This document provides further information on: (\textit{i}) the UMI workflow; {(\textit{ii}) the RPSF and the common midpoint basis}; (\textit{iii}) the comparison between iterative time reversal and phase reversal; (\textit{iv}) the bias of the $\mathbf{T}-$matrix estimator; (\textit{v}) the comparison between a multi-scale and local analysis of wave distortions; (\textit{vi}) the impact of the confocal filter; (\textit{vii}) the effect of an incompleteness of the illumination basis. 

\section{Workflow}
\begin{figure}[ht!]
\includegraphics[width=\textwidth]{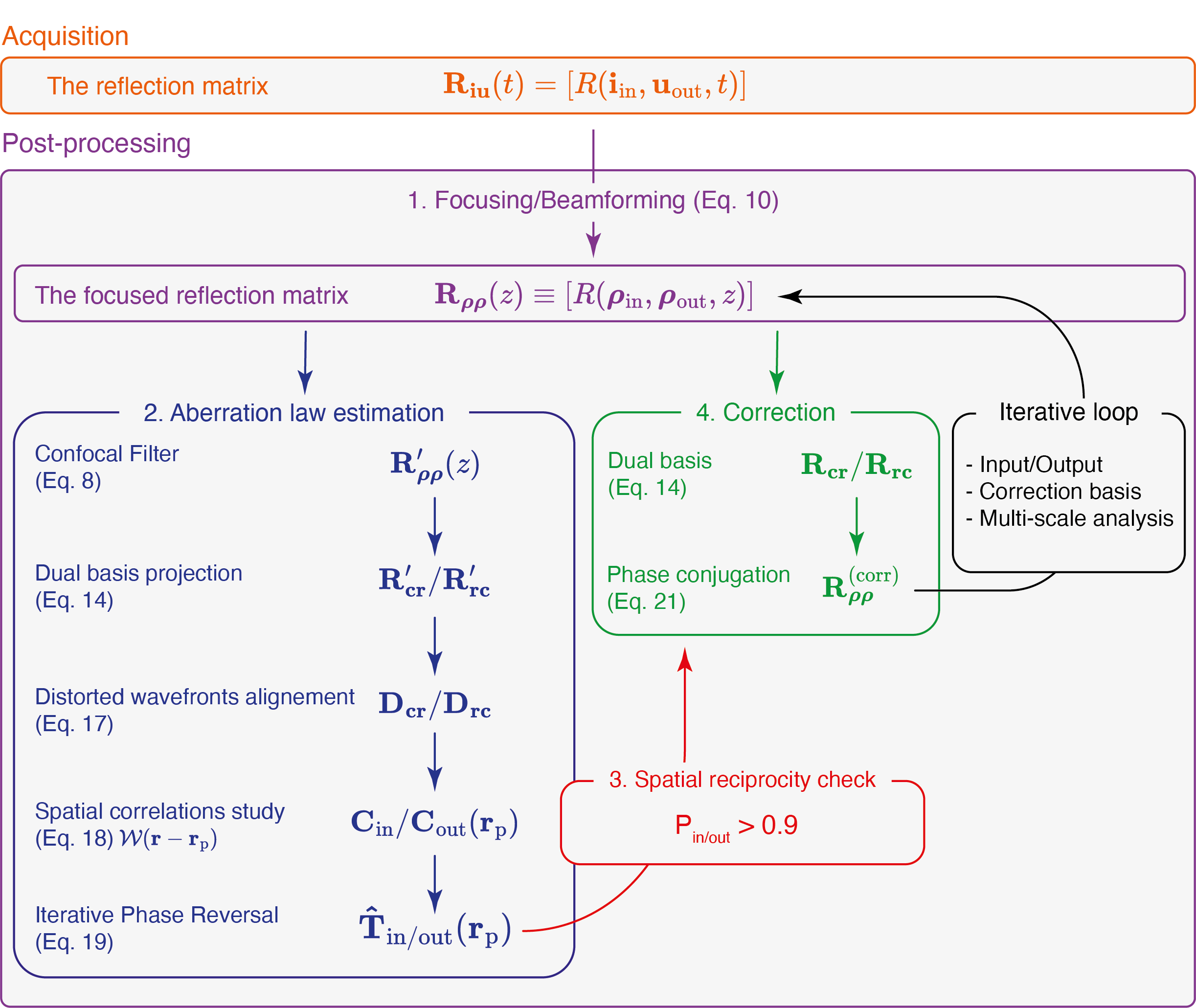}
 \caption{\textbf{Flowchart of the UMI process}. %bis{changer le numéro des équations pour être cohérent avec le texte principal quand il sera fini}-> OK
 }
 \label{S1_Flowchart}
\end{figure}

Supplementary Figure~\ref{S1_Flowchart} shows a workflow that sums up the different steps of the UMI procedure performed in the accompanying paper.

\section{RPSF and common midpoint}
\begin{figure}[tb!]
\includegraphics[width=\textwidth]{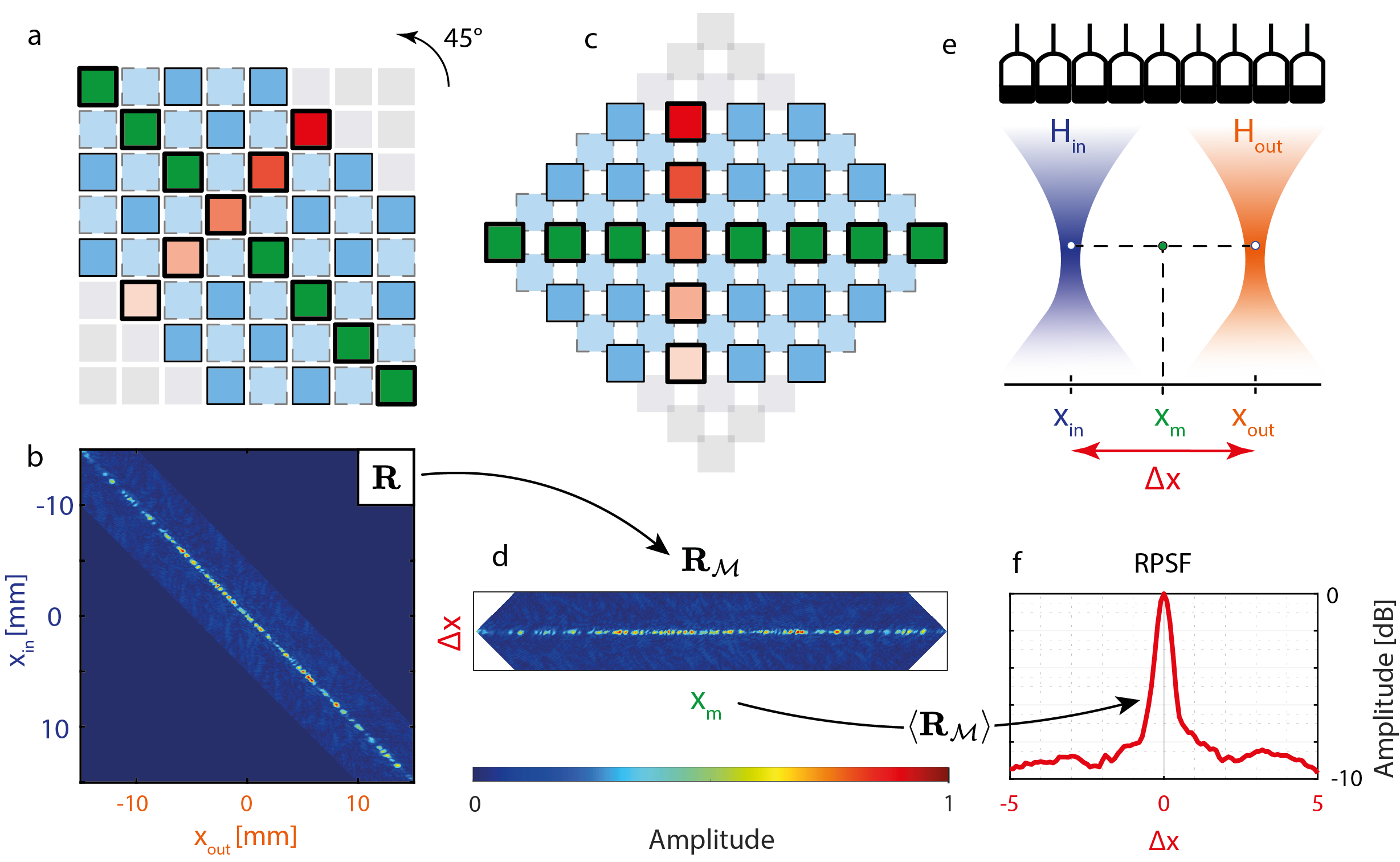}
\caption{{\textbf{Common midpoint representation.} In 2D ultrasound imaging with linear or convex probes, the change from a (\textbf{a},\textbf{b}) conventional to a (\textbf{c},\textbf{d}) common midpoint representation corresponds to a 45$^\circ$ rotation of the focused reflection matrix. Panels a and c show a schematic representation of such a transformation, while panels b and d show experimental ultrasound data in speckle of an ultrasound phantom with a linear probe. Note that the change from the focused to the common midpoint representation implies two new sampling grids, represented by solid and dashed lines. (\textbf{e}) Schematic representation of the position of the input ($x_\textrm{in}$) and output ($x_\textrm{out}$) focal spots, spaced by $\Delta x$ and their common midpoint $x_\textrm{m}$. (\textbf{f}) Extracted RPSF by spatial averaging over all midpoints $x_\textrm{m}$ at depth $z=30$ mm.}}
    \label{cmpfig}
\end{figure}

{To probe the local focusing quality, the reflection point spread function (RPSF) can be investigated. Its extraction from the focused reflection matrix, $\mathbf{R}_{\rhog\rhog}(z)=[R(\rhoin,\rhoout,z)]$, consists in the following change of variable to project the data into a common midpoint basis: 
\begin{equation}
\underbrace{\begin{bmatrix}
\rhog_{\textrm{in}}\\
\rhog_{\textrm{out}}\\
z
\end{bmatrix}}_{\textrm{Focused}}
\rightarrow
\underbrace{\begin{bmatrix}\Delta \rhog\\ \rhog_m\\z\end{bmatrix}=\begin{bmatrix} \rhog_{\textrm{out}} -\rhog_{\textrm{in}}\\\frac{\rhog_{\textrm{in}}+\rhog_{\textrm{out}}}{2}\\z\end{bmatrix}
}_{\textrm{Common midpoint}}.
\end{equation}
This operation is described schematically in Supplementary Figure~\ref{cmpfig} for the simple case of 2D imaging with a linear array of transducers. It consists in extracting each antidiagonal of the focused reflection matrix $\R_{xx}(z)$ (red boxes in Supplementary Figure~\ref{cmpfig}a), corresponding to a  matrix rotation by $45^\circ$. In this representation, $x_\textrm{m}=(x_{\textrm{in}}+x_{\textrm{out}})/2$ is the common midpoint between the input and output focal spot, with the two separated by a distance $\Delta x=x_{\textrm{out}}-x_{\textrm{in}}$. These considerations can be extended to 3D imaging, so that the transverse coordinate, previously $x$, now becomes $\rhog=(x,y)$.}

\section{Correlation matrix of wave distortions}

In the accompanying paper, an iterative phase reversal (IPR) process and a multi-scale analysis of $\mathbf{D}$ have been implemented to retrieve the $\mathbf{T}-$matrix. In the following, we provide a theoretical framework to justify this process, outline its limits and conditions of success. For sake of lighter notation, the dependence over $\rgp$ will be omitted in the following.  

At each step of the aberration correction process, a local correlation matrix of $\mathbf{D}$ is computed. The UMI process assumes the convergence of the correlation matrix $\mathbf{C}$ towards its ensemble average $\left \langle \mathbf{C} \right \rangle$, the so-called covariance matrix~\cite{lambert_distortion_2020,lambert_ultrasound_2022}. In fact, this convergence is never fully realized and $\mathbf{C}$ should be decomposed as the sum of this covariance matrix $\left \langle \mathbf{C}\right \rangle $ and a perturbation term $\delta \mathbf{C}$:
\begin{equation}
\label{C}
   \mathbf{C}= \left \langle \mathbf{C}\right \rangle  +  \delta \mathbf{C}.
\end{equation}
The intensity of the perturbation term scales as the inverse of the number $N_{\mathcal{W}}=(w_\rho^2 w_z)/(\delta \rho_0^2 \delta z_0) $ of resolution cells in each sub-region~\cite{robert_greens_2008,lambert_distortion_2020,lambert_ultrasound_2022}:
\begin{equation}
\label{perturbation}
    \left \langle \left |\delta C(\mathbf{c},\mathbf{c}',\rgp)\right |^2 \right \rangle = \frac{ \left \langle \left | C(\mathbf{c},\mathbf{c}',\rgp)\right |^2 \right \rangle}{N_{\mathcal{W}}}
\end{equation}
This perturbation term can thus be reduced by increasing the size of the spatial window $\mathcal{W}$, but at the cost of a resolution loss. In the following, we express theoretically the bias induced by this perturbation term on the estimation of $\mathbf{T}$-matrices. In particular, we will show how it scales with $N_{\mathcal{W}}$ in each spatial window $\mathcal{W} $ and the focusing quality. To that aim, we will consider the output correlation matrix $\mathbf{C}_{{\textrm{out}}}$ but a similar demonstration can be performed at input. 

\section{Covariance matrix: Synthesis of a virtual guide star}

Under assumptions of local isoplanicity in each spatial window and random reflectivity, the covariance matrix can be expressed as follows~\cite{lambert_distortion_2020}:
\begin{equation}
\label{rhoaveeq}
\left \langle \mathbf{C}_{{\textrm{out}}} \right \rangle  =  \mathbf{T}_\textrm{out} \times  \mathbf{C}_H \times \mathbf{T}_\textrm{out}^{\dag},
\end{equation}
or in terms of matrix coefficients,
\begin{equation}
\label{rhoaveeq2}
\left \langle \mathbf{C} (\mathbf{c},\mathbf{c}') \right \rangle =  {T}_\textrm{out} (\mathbf{c})  {T}_\textrm{out}^* (\mathbf{c}')  \underbrace{\int d\bm{\rho} |H_\textrm{in}(\bm{\rho})|^2 e^{-i2\pi \frac{(\mathbf{c}-\mathbf{c}').\mathbf{\rho}}{\lambda z_p}}}_{=C_H(\mathbf{c},\mathbf{c}')}.
\end{equation}
$\mathbf{C}_H$ is a reference correlation matrix associated with a virtual reflector whose scattering distribution corresponds to the input focal spot intensity $|H_\textrm{in}(\bm{\rho})|^2$. This scatterer plays the role of virtual guide star in the UMI process (Fig.~1k of the accompanying paper).  
%\section{Estimation of the $\mathbf{T}-$matrix}

\section{Comparison between iterative time reversal and phase reversal}

In previous works on 2D UMI~\cite{lambert_distortion_2020,lambert_ultrasound_2022}, the $\mathbf{T}$-matrix was estimated by performing a singular value decomposition of $\mathbf{D}_{\mathbf{rc}} $:
  \begin{equation}
\label{eq:svd}
\mathbf{D}_{\mathbf{rc}} =\mathbf{V}_\textrm{in}^{\dag}\times \mathbf{\Sigma}\times \mathbf{U}_\textrm{out} ,
\end{equation}
or, equivalently, the eigenvalue decomposition of $\mathbf{C}_{{\textrm{out}}} $:
  \begin{equation}
\label{eq:evd}
\mathbf{C}_{{\textrm{out}}} =\mathbf{U}_\textrm{out}^{\dag} \times \mathbf{\Sigma}^2\times \mathbf{U}_\textrm{out}.
\end{equation}
$\mathbf{\Sigma}$ is a diagonal matrix containing the singular value{s} $\sigma_i$ in descending order: $\sigma_1>\sigma_2>..>\sigma_N$. $\mathbf{U}_\textrm{out}$ and $\mathbf{V}_\textrm{in}$ are unitary matrices that contain the orthonormal set of output and input eigenvectors, $\mathbf{U}^{(i)}_\textrm{out}=[U_\textrm{out}^{(i)}({\mathbf{c}})]$ and $\mathbf{V}^{(i)}_\textrm{in}=[V_\textrm{in}^{(i)}(\rg)]$.

\begin{figure}[ht!]
\includegraphics[scale=1.4]{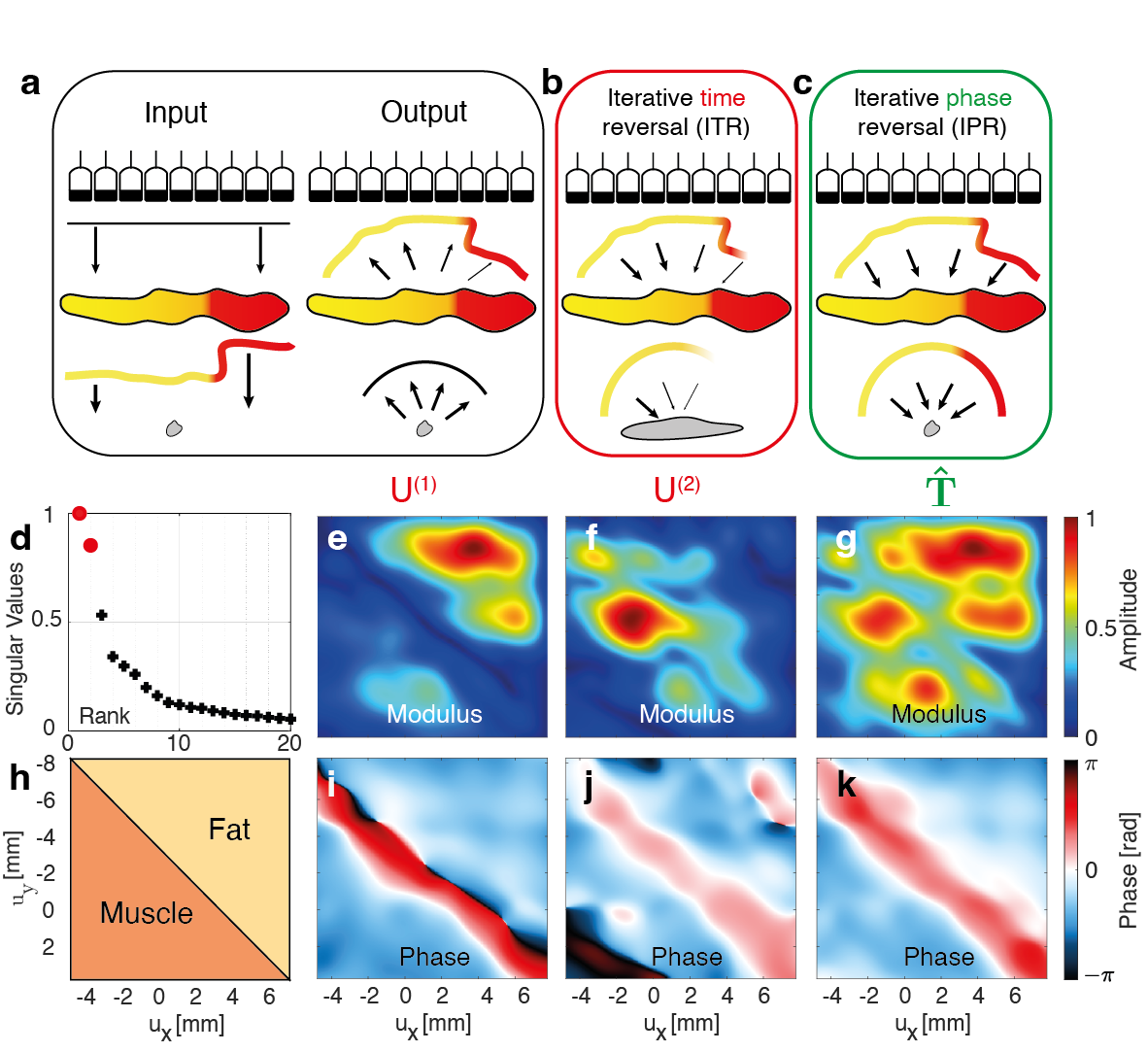}
  \caption{\textbf{Iterative Time Reversal vs. Iterative Phase Reversal.} (\textbf{a}) The first step of ITR and IPR corresponds to the following fictitious experiment: Insonifying the medium by an arbitrary wave-front (here a plane wave) using an array of transducers and recording the reflected wave-front with the same probe. (\textbf{b}) The ITR process consists in time-reversing this wave-front in post-processing and sending it back into the medium, recording again the reflected wave-front, and so on. (\textbf{c}) The IPR process is similar but normalizes the amplitude of the time-reversed wavefront at every iteration. (\textbf{d}) Singular value distribution of $\mathbf{D}_{\mathbf{rc}}$ for a box $\mathcal{W}$ of dimension $\mathbf{w}=(w_x,w_y,w_z)=(2,-5,2)$mm centered around point $\rgp=(3,-5.5,23)$ mm. (\textbf{e},\textbf{f}) Modulus of the two first eigenvectors $\mathbf{U}^{(i)}_\textrm{out}$. (\textbf{g}) Modulus of the vector ${\mathbf{C}_{\textrm{out}}}\times \hat{\mathbf{T}}_\textrm{out}$.  (\textbf{h}) Delimitation of muscle and fat over the probe surface. (\textbf{i},\textbf{j},\textbf{k}) Phase of $\mathbf{U}^{(1)}_\textrm{out}$, $\mathbf{U}^{(2)}_\textrm{out}$ and $\hat{\mathbf{T}}_\textrm{out}$.}%{à quoi correspond le code couleur du front d'onde ? Je pensais que c'était l'amplitude, mais sur le pannel b l'amplitude semble être exprimée en transparence.}} -> vitesse du son ans l'aberrateur
  \label{S2_IPRvsSVD}
\end{figure}

The reason of this eigenvalue decomposition can be intuitively understood by considering the asymptotic case of a point-like input focusing beam. In this ideal case, Eq.~\ref{eq:evd} becomes $ C_{{\textrm{out}}}( \mathbf{c}, \mathbf{c}' ) =  T_\textrm{out}(\mathbf{c}) T_\textrm{out}^*(\mathbf{c}')$. 
$\mathbf{D}_{\mathbf{rc}}$ is then of rank $1$ -- the %. The 
first output singular vector $\mathbf{U}^{(1)}_\textrm{out}$ yields the aberration transmittance $\mathbf{T}_\textrm{out}$.

However, in reality, the input PSF $H_\textrm{in}$ is of course far from being point-like. The spectrum of $\mathbf{D}_{\mathbf{rc}}$ displays a continuum of singular values [Supplementary Figure~\ref{S2_IPRvsSVD}d]. The effective rank of $\mathbf{C}_{{\textrm{out}}}$ is shown to scale as the number of resolution cells covered by the input PSF $H_\textrm{in}$~\cite{lambert_ultrasound_2022}:
\begin{equation}
\label{rank}
    M_\delta \sim  (\delta \rho_\textrm{in} /\delta \rho_0)^2.
\end{equation}
with $\delta \rho_\textrm{in}$ the spatial extension of {$H_\textrm{in}$}. The amplitude of the corresponding eigenvectors $\mathbf{U}^{(i)}_\textrm{out}$ depends on the exact shape of the virtual guide star, that is to say, on aberrations induced by the incident wave-front. 

Supplementary Figures~\ref{S2_IPRvsSVD}e and f show the modulus of two first eigenvectors, $\mathbf{U}^{(1)}_\textrm{out}$ and $\mathbf{U}^{(2)}_\textrm{out}$. They clearly show a complementary feature. While $\mathbf{U}^{(1)}_\textrm{out}$ is associated with the fat layer, $\mathbf{U}^{(2)}_\textrm{out}$ maps onto the muscle part of the pork chop [Supplementary Figure~\ref{S2_IPRvsSVD}h]. This result can be understood by the discontinuity of the speed-of-sound between the muscle and fat parts of the pork chop that breaks the spatial invariance and isoplanicity. As a consequence, the SVD process tends to converge onto eigenstates associated with the most isoplanatic components of $\mathbf{D}_{\mathbf{r}\mathbf{c}}$. 

This property is not satisfactory in the present case since each eigenvector only covers a part of the probe aperture. In other words, the phases of $\mathbf{U}^{(1)}$ [Supplementary Figure~\ref{S2_IPRvsSVD}i]  and $\mathbf{U}^{(2)}$ [Supplementary Figure~\ref{S2_IPRvsSVD}j] are only satisfying estimators of $\mathbf{T}$ over some parts of the probe. Therefore, they cannot independently lead to an aberration compensation over the full numerical aperture. 

To circumvent that problem, one can take advantage of the analogy with iterative time reversal (ITR). The eigenvector $\mathbf{U}^{(1)}_\textrm{out}$ can actually be seen as the result of the following fictitious experiment that consists in illuminating the virtual scatterer by an arbitrary wave-front and recording the reflected wave-field [Supplementary Figure~\ref{S2_IPRvsSVD}a]. This wave-field is time-reversed and back-emitted towards the virtual scatterer [Supplementary Figure~\ref{S2_IPRvsSVD}b]. This process can then be iterated many times and each step can be mathematically written as:
\begin{equation}
\label{ITR}
\sigma    \mathbf{W}^{(n+1)}= \mathbf{C}_{{\textrm{out}}}\times \mathbf{W}^{(n)}
\end{equation}
with $\mathbf{W}^{(n)}$, the wave-front at iteration $n$ of the ITR process and $\sigma$, the scatterer reflectivity. ITR is shown to converge towards a time-reversal invariant that is nothing other than the first eigenvector,  $\mathbf{U}^{(1)}_\textrm{out}=\lim\limits_{n \rightarrow +\infty} \mathbf{W}^{(n)}$.

To optimize the estimation of aberrations over the full probe aperture, our idea is to modify the ITR process by still re-emitting a phase-reversed wave-field but with a constant amplitude on each probe element [Supplementary Figure~\ref{S2_IPRvsSVD}c]. In practice, this operation is performed using the following IPR algorithm:
\begin{equation}
  \hat{\mathbf{T}}_\textrm{out}^{(n+1)} = \exp\left[i \, \mathrm{arg}\left \lbrace   \mathbf{C}_{{\textrm{out}}}\times {\hat{\mathbf{T}}}_\textrm{out}^{(n)}  \right \rbrace \right]
\end{equation}
where $\hat{\mathbf{T}}_\textrm{out}^{(n)} $ is the estimator of ${\mathbf{T}}_\textrm{out} $ at the $n^\textrm{th}$ iteration of IPR.  $ \hat{\mathbf{T}}_\textrm{out}^{(0)}$ is an arbitrary wave-front that initiates IPR (typically a plane wave). $\hat{\mathbf{T}}_\textrm{out} = \lim_{n\to\infty} \hat{\mathbf{T}}_\textrm{out}^{(n)}$ is the result of this IPR process.
Unlike ITR, IPR equally addresses each angular component of the imaging process to reach a diffraction-limited resolution. Supplementary Figure~\ref{S2_IPRvsSVD}g illustrates this fact by showing the modulus of $  \mathbf{C}_{{\textrm{out}}}\times \hat{\mathbf{T}}_\textrm{out}$. Compared with $\mathbf{U}^{(1)}_\textrm{out}$ [Supplementary Figure~\ref{S2_IPRvsSVD}e] and $\mathbf{U}^{(2)}_\textrm{out}$ [Supplementary Figure~\ref{S2_IPRvsSVD}f], it clearly shows that the phase-reversed invariant $\hat{\mathbf{T}}_\textrm{out}$ simultaneously addresses each angular component of the aberrated wave-field.  $\hat{\mathbf{T}}_\textrm{out}$ is thus a much better estimator of the ${\mathbf{T}}-$matrix [Supplementary Figure~\ref{S2_IPRvsSVD}k] than the aberration phase laws extracted by the SVD process [Supplementary Figures~\ref{S2_IPRvsSVD}i and j]. 

When applied to the whole field-of-view, the IPR algorithm is mathematically equivalent to the CLASS algorithm developed in optical microscopy~\cite{kang_high-resolution_2017}. However, the IPR algorithm is much more efficient for a local compensation of aberrations. For IPR, the angular resolution $\delta \theta$ of the aberration phase law is only limited by the angular pitch of the plane wave illumination basis or the pitch $p$ of the transducer array in the canonical basis: $\delta \theta_I \sim \lambda /p$. With CLASS, the resolution $\delta \theta_{C}$ of the aberration law is governed by the size of the spatial window $\mathcal{W}$ on which the focused reflection matrix is truncated: $\delta \theta_C \sim z/w_\rho$. It can be particularly detrimental when high-order aberrations and small isopalanatic patches are targeted.

\section{Bias on the $\mathbf{T}-$matrix estimation}

In practice, however, the $\mathbf{T}-$matrix estimator is still impacted by the blurring of the synthesized guide star and the presence of diffusive background and/or noise.  Therefore, the whole process shall be iterated at input and output in order to gradually refine the guide star and reduce the bias on our $\mathbf{T}-$matrix estimator. Moreover, the spatial window $\mathcal{W}$ over which the $\mathbf{C}-$matrix is computed shall be gradually decreased in order to address the high-order aberration components, the latter one being associated with smaller isoplanatic patches. 

To understand the parameters controlling the bias {$\delta {\mathbf{T}}_\textrm{out} $} between $\hat{\mathbf{T}}_\textrm{out}$ and ${\mathbf{T}}_\textrm{out}$, one can express $\hat{\mathbf{T}}_\textrm{out}$ as follows:
\begin{equation}
 \hat{\mathbf{T}}_\textrm{out}=\exp \left ( j \mbox{arg} \left \lbrace  \mathbf{C}_{{\textrm{out}}} \times \hat{\mathbf{T}}_\textrm{out}  \right \rbrace  \right )   =  \frac{ \mathbf{C}_{{\textrm{out}}}  \times \hat{\mathbf{T}}_\textrm{out} }{|| \mathbf{C}_{{\textrm{out}}}  \times \hat{\mathbf{T}}_\textrm{out} ||}
\end{equation}
By injecting Eq.~\ref{C} into the last expression, $\hat{\mathbf{T}}_\textrm{out}$ can be expressed, at first order, as the sum of its expected value ${\mathbf{T}}_\textrm{out}$ and a perturbation term $\delta \hat{\mathbf{T}}_\textrm{out} $:
\begin{equation}
   \hat{\mathbf{T}}_\textrm{out}=\underbrace{\frac{ \langle \mathbf{C}_{{\textrm{out}}}\rangle \times {\mathbf{T}}_\textrm{out}  }{|| \langle \mathbf{C}_{{\textrm{out}}} \rangle \times {\mathbf{T}}_\textrm{out} ||}}_{= \mathbf{T}_\textrm{out}} + \underbrace{\frac{  \delta \mathbf{C}_{{\textrm{out}}} \times {\mathbf{T}}_\textrm{out}}{|| \langle \mathbf{C}_{{\textrm{out}}} \rangle \times {\mathbf{T}}_\textrm{out}  ||}}_{\simeq \delta \hat{\mathbf{T}}_\textrm{out}}.
\end{equation}
The bias intensity can be expressed as follows:
\begin{equation}
\label{bias}
    {\lvert \delta {\mathbf{T}}_\textrm{out} \rvert^2}=\frac{ {\mathbf{T}}_\textrm{out}^{\dag} \times   \delta \mathbf{C}_{{\textrm{out}}}^{\dag}\times \delta \mathbf{C}_{{\textrm{out}}} \times   {\mathbf{T}_\textrm{out}}}{ {\mathbf{T}}_\textrm{out}^{\dag}\times \langle \mathbf{C}_{{\textrm{out}}}\rangle^{\dag} \times \langle \mathbf{C}_{{\textrm{out}}} \rangle \times   {\mathbf{T}}_\textrm{out}}
\end{equation}
Using Eq.~\ref{perturbation}, the numerator of the last equation can be expressed as follows:
 \begin{equation}
  {\mathbf{T}}_\textrm{out}^{\dag} \times   \delta \mathbf{C}_{{\textrm{out}}}^{\dag}\times \delta \mathbf{C}_{{\textrm{out}}} \times   {\mathbf{T}_\textrm{out}} = N_{u}^2  \langle | {\delta C}(\mathbf{c},\mathbf{c}')|^2\rangle = N_{u}^2 {|  {C} (\mathbf{c},\mathbf{c})|^2 } /N_{\mathcal{W}}  .
 \end{equation}
 with $N_{u}$ the number of transducers.% and $\ell_u$ the coherence length of the aberrating layer seen from the transducer basis.{$\ell_u$ n'apparait pas dans l'equation, $N_{\mathcal{W}}$??} -> oui il rallait le virer
 %Injecting Eq.~\ref{rhoaveeq2} into the last equation leads to the following expression for the numerator of Eq.~\ref{bias}:
  %\begin{equation}
  %{\mathbf{T}}_\textrm{out}^{\dag} \times   \delta \mathbf{C}_{\bm{uu}}^{\dag}\times \delta \mathbf{C}_{\bm{uu}} \times   {\mathbf{T}_\textrm{out}}= N^2 \ell_u^2  \left | T_\textrm{in} \stackrel{\mathbf{u}_\textrm{in}}{\circledast} T_\textrm{in} (\mathbf{0}) \right |^2  /N_\textrm{w} .
 %\end{equation}
 
The denominator of Eq.~\ref{bias} can be expressed as follows:
\begin{equation}
{\mathbf{T}}_\textrm{out}^{\dag} \times \langle \mathbf{C}_{{\textrm{out}}}\rangle^{\dag} \times \langle \mathbf{C}_{{\textrm{out}}} \rangle \times   {\mathbf{T}}_\textrm{out}= M^2\left | \sum_{\mathbf{c}}  T_\textrm{in} \stackrel{\mathbf{c}}{\circledast} T_\textrm{in} (\mathbf{c})  \right |^2
\end{equation}
The bias intensity is thus given by:
\begin{equation}
   {\lvert \delta {{T}}_\textrm{out} (\mathbf{c}) \rvert^2} = \frac{\left | T_\textrm{in} \stackrel{\mathbf{c}}{\circledast} T_\textrm{in} (\mathbf{0}) \right |^2}{N_{\mathcal{W}}\left | \sum_{\mathbf{c}}  T_\textrm{in} \stackrel{\mathbf{c}}{\circledast} T_\textrm{in} (\mathbf{c})  \right |^2} 
\end{equation}
In the last expression, we recognize the ratio between the coherent intensity (energy deposited exactly at focus) and the mean incoherent input intensity. This quantity is known as the coherence factor in ultrasound imaging~\cite{mallart_adaptive_1994,robert_greens_2008}:
\begin{equation}
    \mathcal{C}_\textrm{in}= \frac{  \sum_{\mathbf{c}}  T_\textrm{in} \stackrel{\mathbf{c}}{\circledast} T_\textrm{in} (\mathbf{c}) }{  T_\textrm{in} \stackrel{\mathbf{c}}{\circledast} T_\textrm{in} (\mathbf{0})  } =\frac{|H_\textrm{in} (\bm{\rho}=\mathbf{0})|^2}{   \Delta \rho_\textrm{max}^{-2}\int d\bm{\rho} |H_\textrm{in} (\bm{\rho})|^2} 
\end{equation}
In the speckle regime and for a 2D probe, the coherence factor $\mathcal{C}$ ranges from 0,
for strong aberrations and/or multiple scattering background, to $4/9$ in the ideal case~\cite{Silverstein2001}. The bias intensity can thus be rewritten as:
\begin{equation}
   {| \delta {{T}}_\textrm{out} (\mathbf{c}) |^2} = \frac{1}{\mathcal{C}^2_\textrm{in} N_{\mathcal{W}}} 
\end{equation}
This last expression justifies the multi-scale analysis proposed in the accompanying paper. A gradual increase of the focusing quality, quantified by $\mathcal{C}$, is required to address smaller spatial windows that scale as $N_{\mathcal{W}}$. Following this scheme, the bias made of our $\mathbf{T}-$matrix estimator can be minimized. 

\section{{Probing the bias intensity with spatial reciprocity}}

{In the accompanying paper, we use the scalar product $P_{\textrm{in/out}}$ between input and output aberration phase laws to monitor the bias $ {| \delta {{T}} |^2} $ of our $\mathbf{T}-$matrix estimator. Here we demonstrate the link between both quantities. To do so, the estimator can be written as:
\begin{equation}
\label{eq1bis}
\hat{{T}}(\mathbf{c},\mathbf{r}_{\textrm{p}})=\exp\left [ j \left \lbrace \phi(\mathbf{c},\mathbf{r}_{\textrm{p}}) +\delta \phi(\mathbf{c},\mathbf{r}_{\textrm{p}})  \right \rbrace \right ]
\end{equation}
with ${{T}}(\mathbf{c},\mathbf{r}_{\textrm{p}})=\exp\left [ j  \phi(\mathbf{c},\mathbf{r}_{\textrm{p}})  \right ]$ and $\delta \phi(\mathbf{c},\mathbf{r}_{\textrm{p}})$ the phase error of the estimator.}

{On the one hand, the  bias intensity can be rewritten using Eq.~\ref{eq1} as follows:
\begin{equation}
\label{eq2}
|\delta {T}(\mathbf{c},\mathbf{r}_{\textrm{p}})|^2= \left|1-\exp[j\delta \phi (\mathbf{c},\mathbf{r}_{\textrm{p}})]\right |^2 = 4 \sin^2 \left [ \frac{\delta \phi (\mathbf{c},\mathbf{r}_{\textrm{p}}))}{2} \right ] \stackrel{\delta \phi <<1}{\sim} \left [\delta \phi (\mathbf{c},\mathbf{r}_{\textrm{p}}) \right ]^2    
\end{equation}
On the other hand, the scalar product $P_{\textrm{in}/\textrm{out}}$ is given by
\begin{equation}
P_{\textrm{in}/\textrm{out}}=N_c^{-1} \sum_{\mathbf{c}} \exp\left [ j \left \lbrace \delta \phi_\textrm{in}(\mathbf{c},\mathbf{r}_{\textrm{p}})  - \delta \phi_\textrm{out}(\mathbf{c},\mathbf{r}_{\textrm{p}})  \right \rbrace \right ]
\end{equation}
In the previous equation, the sum over $\mathbf{c}$ can be replaced by an ensemble average since $N_c=N_u>>1$:
\begin{equation}
P_{\textrm{in}/\textrm{out}}= \left \langle \exp\left [ j \left \lbrace \delta \phi_\textrm{in}(\mathbf{c},\mathbf{r}_{\textrm{p}})  - \delta \phi_\textrm{out}(\mathbf{c},\mathbf{r}_{\textrm{p}})  \right \rbrace   \right ]\right \rangle. 
\end{equation}
Assuming a small phase error ($\delta \phi_\textrm{in/out}<<1$), the last equation can be rewritten as follows
\begin{equation}
P_{\textrm{in}/\textrm{out}} \simeq 1 + j \left \langle \delta \phi_\textrm{in}(\mathbf{c},\mathbf{r}_{\textrm{p}}) - \delta \phi_\textrm{out}(\mathbf{c},\mathbf{r}_{\textrm{p}})  \right \rangle -\frac{\langle [\delta \phi_\textrm{in}(\mathbf{c},\mathbf{r}_{\textrm{p}})-\delta \phi_\textrm{out}(\mathbf{c},\mathbf{r}_{\textrm{p}})]^2 \rangle }{2}.  
\end{equation}
Since $\langle \delta \phi_\textrm{in/out} \rangle=0 $ and $\langle \delta \phi_\textrm{in} \delta \phi_\textrm{out}  \rangle =0$, the last expression simplifies into
\begin{equation}
P_{\textrm{in}/\textrm{out}} \simeq 1 - \frac{\langle |\delta \phi_\textrm{in}(\mathbf{c},\mathbf{r}_{\textrm{p}})|^2}{2} - \frac{\langle |\delta \phi_\textrm{out}(\mathbf{c},\mathbf{r}_{\textrm{p}})|^2 \rangle}{2}.  
\end{equation}
Assuming an equivalent phase error at input and output ($\langle |\delta \phi_\textrm{in}(\mathbf{c},\mathbf{r}_{\textrm{p}})|^2 \rangle=\langle |\delta \phi_\textrm{out}(\mathbf{c},\mathbf{r}_{\textrm{p}})|^2 \rangle$) finally leads to:
\begin{equation}
P_{\textrm{in}/\textrm{out}} \simeq 1 - \langle |\delta \phi(\mathbf{c},\mathbf{r}_{\textrm{p}})|^2  \rangle .  
\end{equation}
Combining the latter expression with Eq.~\ref{eq2} leads to the final result:
\begin{equation}
P_{\textrm{in}/\textrm{out}} \simeq 1 - \left \langle |\delta {T}(\mathbf{c},\mathbf{r}_{\textrm{p}})|^2 \right\rangle.  
\end{equation}
$P_{\textrm{in}/\textrm{out}}$ is thus a relevant quantity to estimate the bias intensity (see Fig.~3b of the accompanying paper).}

\section{Multi-scale analysis of wave distortions}
\begin{figure}[ht!]
\includegraphics[width=\textwidth]{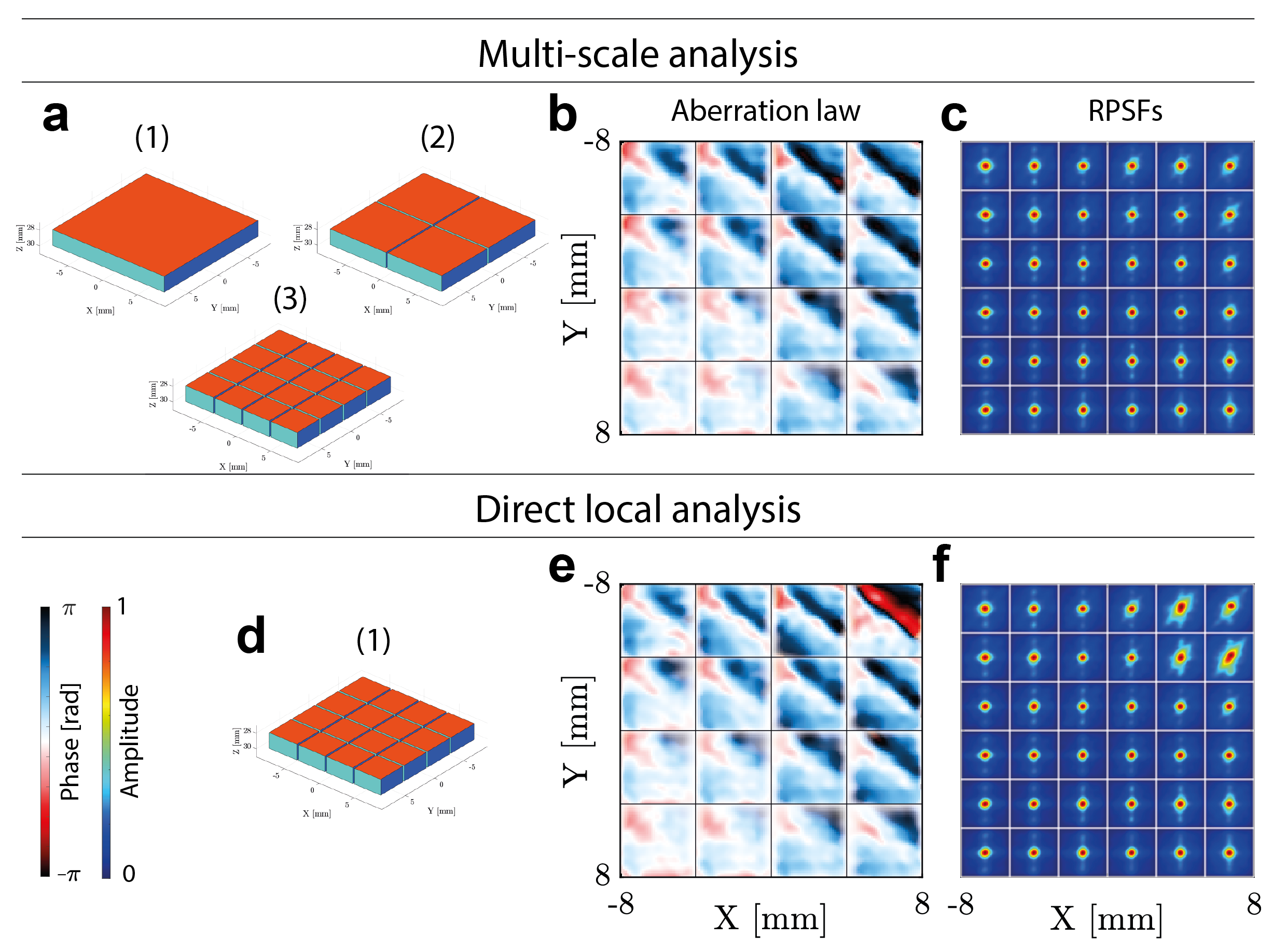}
 \caption{\textbf{Multi-scale \textit{versus} direct local analysis of wave distortions}  (pork chop experiment, $z=29$ mm). (\textbf{a}) Representation of the spatial windows used at each step of UMI (see Tab.~III of the accompanying paper). (\textbf{b}) Aberration phase laws ($\hat{\mathbf{T}}$) extracted by a multi-scale analysis. (\textbf{c}) RPSFs after multi-scale aberration compensation. (\textbf{d}) Representation of the spatial windows used for a direct local compensation of wave distortions. (\textbf{e}) Aberration phase laws ($\hat{\mathbf{T}}$) extracted by a local analysis of $\mathbf{D}$. (\textbf{f}) RPSFs after local aberration compensation.}
 \label{S3_ProgressiveVersusDirect}
\end{figure}
Supplementary Figure~\ref{S3_ProgressiveVersusDirect} demonstrates the benefit of a multi-scale analysis of wave distortions with a gradual decrease of spatial windows $\mathcal{W}$ at each step of the UMI process [Supplementary Figure~\ref{S3_ProgressiveVersusDirect}a]. To that aim, this aberration correction scheme is compared with a direct estimation of the $\mathbf{T}-$matrix over the smallest patches $\mathcal{W}$ [Supplementary Figure~\ref{S3_ProgressiveVersusDirect}d]. The estimated transmission matrices $\hat{\mathbf{T}}$ differ in both cases (see comparison between Supplementary Figures~\ref{S3_ProgressiveVersusDirect}b and e) especially in the fat layer. The RPSFs obtained after phase conjugation of $\hat{\mathbf{T}}$ demonstrate the benefit of the multi-scale analysis [Supplementary Figure~\ref{S3_ProgressiveVersusDirect}c] compared with a direct local investigation of wave distortions [Supplementary Figure~\ref{S3_ProgressiveVersusDirect}f].  The fat area is actually the most aberrated in the field-of-view (see initial RPSFs displayed by Fig.~2b of the accompanying paper). The initial coherence factor $\mathcal{C}$ is thus much smaller in this area, which induces a strong bias on $\mathbf{T}$ when wave distortions are investigated over a reduced isoplanatic patch. On the contrary, a multi-scale analysis enables a gradual enhancement of this coherence factor in this area and finally leads to an unbiased estimation of $\mathbf{T}$.   

\begin{figure}[ht!]
\includegraphics[width=0.6\textwidth]{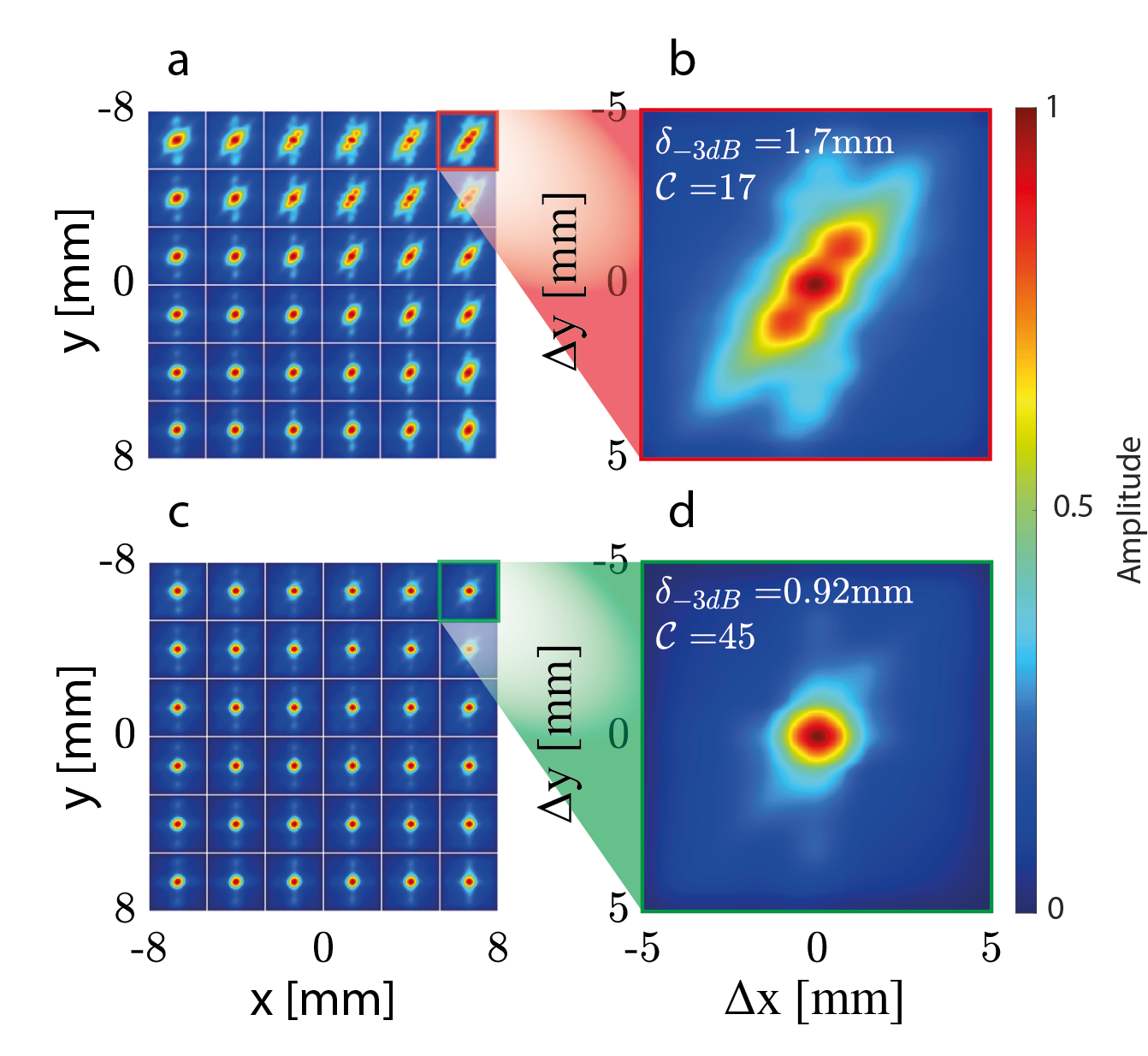}
 \caption{\textbf{Contrast \& resolution enhancement in the pork chop experiment.} (\textbf{a}) Maps of local RPSF ($z=29$ mm). (\textbf{b}) Local RPSF on the top right of the field-of-view. (\textbf{c}) Map of RPSF after the UMI process. (\textbf{d}) Corrected RPSF on the top right of the field-of-view. The resolution is evaluated at $-3$dB (see Methods in the accompanying paper). The contrast {$\mathcal{F}$} is the ratio between the confocal peak and the multiple scattering/noise background  (see also Methods).}
 \label{S4_enhancementCDP}
\end{figure}

Supplementary Figure~\ref{S4_enhancementCDP} shows the performance of UMI by comparing the RPSFs before and after aberration compensation. In the most aberrated area (top right of the field-of-view), the resolution is improved by almost a factor two, while the contrast is increased by 4.2 dB.   

\begin{figure}[ht!]
\includegraphics[width=\textwidth]{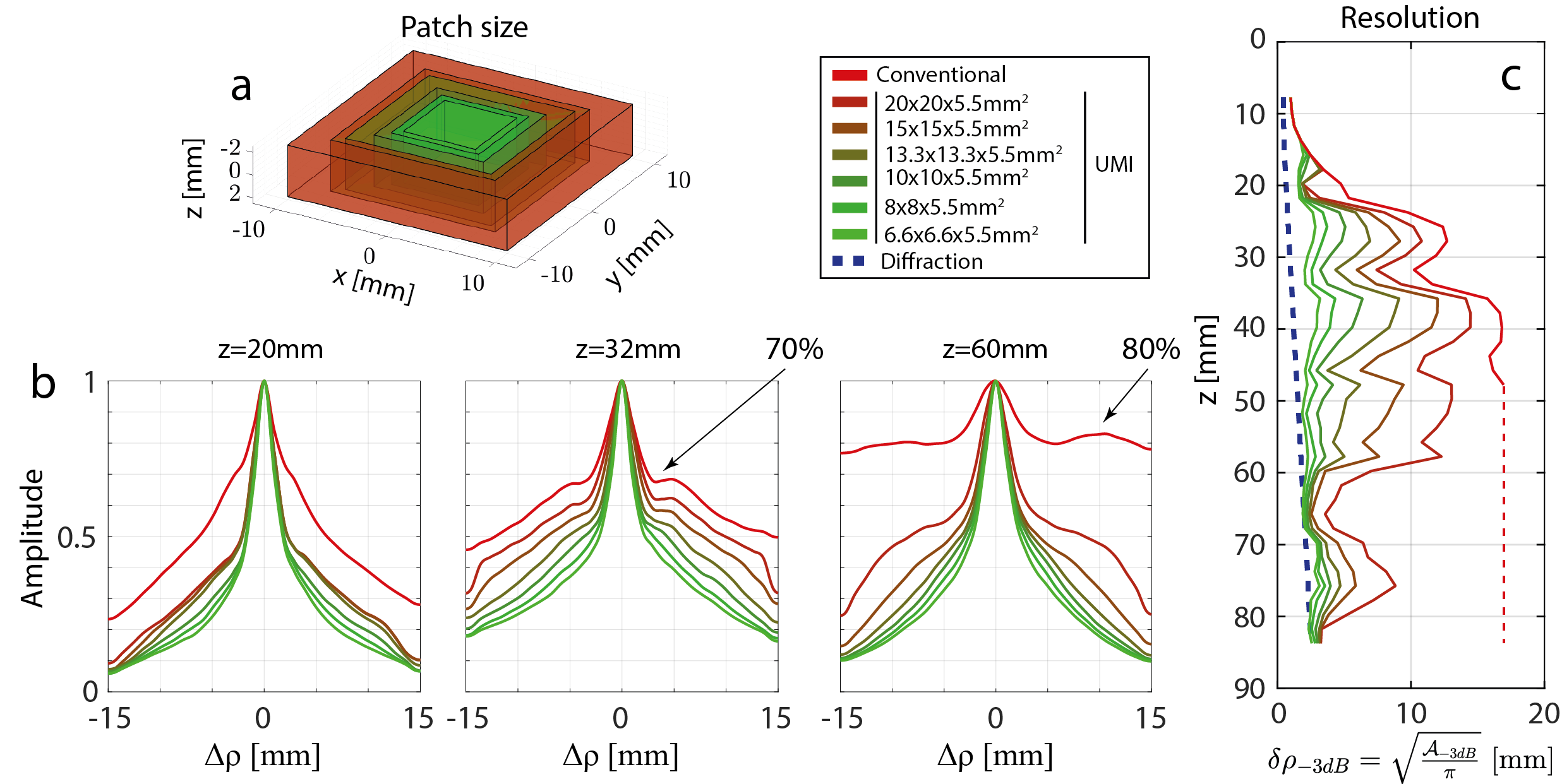}
 \caption{\textbf{Multi-scale compensation of wave distortions in the head phantom.} (\textbf{a}) Successive patches used to perform a multi-scale analysis of wave distortions. (\textbf{b}) Radial profile of the RPSF amplitude at each step for three different depths (From left to right: $z=20$, $z=32$ and $z=60$ mm). (\textbf{c}) Resolution as a function of depth at each step of correction (from red to green). At large depth (red dashed line), initial resolution can not be extracted as the incoherent background is larger than 1/2 as shown in panel (\textbf{b}) for $z=60$ mm.}
 \label{S5_MultiscaleHeadPhantom}
\end{figure}

Supplementary Figure~\ref{S5_MultiscaleHeadPhantom} shows the evolution of the RPSF during the UMI process applied to the head phantom experiment. A gradual enhancement of the focusing process is observed at each step of UMI, which enables an estimation of the $\mathbf{T}-$matrix at a higher resolution.  

\section{Confocal filter}
\begin{figure}[ht!]
\includegraphics[width=\textwidth]{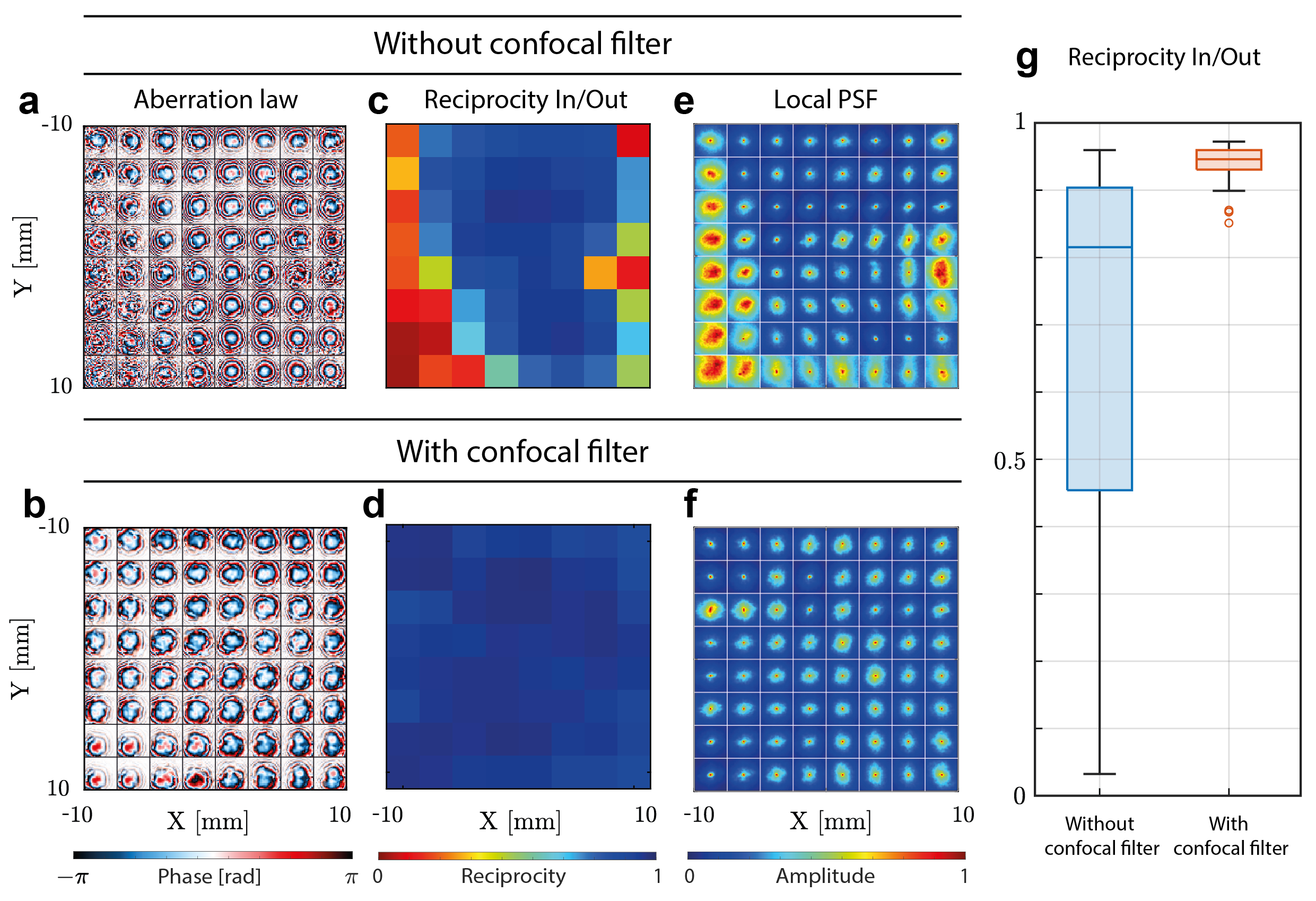}
 \caption{\textbf{Confocal filter in transcranial imaging.} (\textbf{a},\textbf{b}) Output aberration phase laws ($\hat{\mathbf{T}}_\textrm{out}$) extracted without and with a confocal filter. (\textbf{c},\textbf{d}) Normalised scalar products {$P_{\textrm{in/out}}$} without and with a confocal filter, respectively. (\textbf{e},\textbf{f}) RPSFs obtained with UMI without and with a confocal filter. (\textbf{g}) Box plot corresponding to the panels (\textbf{c},\textbf{d}). Experimental data shown in this figure correspond to the head phantom experiment described in the accompanying paper ($z=50mm$). }%{Qu'est ce que tu veux dire par Reciprocity ? C'est le biais décrit plus haut ? Tu parles aussi de corrélation dans le corps du texte} {ce que j'appelle reciprocity c'est le produit scalaire entrée versus sortie : $N_u^{-1}\hat{\mathbf{T}}_\textrm{in}^{\dag}\hat{\mathbf{T}}_\textrm{out}$}}
 \label{S6_ConfocalFilter_and_Reciprocity}
\end{figure}

Supplementary Figure~\ref{S6_ConfocalFilter_and_Reciprocity} shows the effect of the confocal filter on the ${\mathbf{T}}-$matrix estimation. The output aberration phase laws contained in $\hat{\mathbf{T}}_{\textrm{out}}$ look much more noisy in absence of an adaptive confocal filter (see the comparison between Supplementary Figures~\ref{S6_ConfocalFilter_and_Reciprocity}a and b). As shown by the scalar product between input and output aberration phase laws [Supplementary Figure~\ref{S6_ConfocalFilter_and_Reciprocity}c], this ``noise'' comes from the imperfect convergence of $\hat{\mathbf{T}}$ towards ${\mathbf{T}}$. Without any confocal filter, multiple scattering drastically reduces the coherence factor and induces a strong bias on estimation of ${\mathbf{T}}$ (see Supplementary Section S5). On the contrary, the adaptive confocal filter enables an enhancement of this coherence factor $\mathcal{C}$ to ensure a satisfactory estimation of ${\mathbf{T}}$. The high degree of correlation between $\hat{\mathbf{T}}_\textrm{in}$ and $\hat{\mathbf{T}}_\textrm{out}$ proves this last assertion [Supplementary Figure~\ref{S6_ConfocalFilter_and_Reciprocity}d]. The effect of the confocal filter is also particularly obvious when looking at the RPSF obtained at the end of the UMI process. While a strong incoherent background subsists on the lateral parts of the field-of-view when no confocal filter is applied [Supplementary Figure~\ref{S6_ConfocalFilter_and_Reciprocity}e], a homogeneous focusing quality is obtained with the confocal filter [Supplementary Figure~\ref{S6_ConfocalFilter_and_Reciprocity}f]. 

\section{Illumination basis}

\begin{figure}[ht!]
\includegraphics[width=\textwidth]{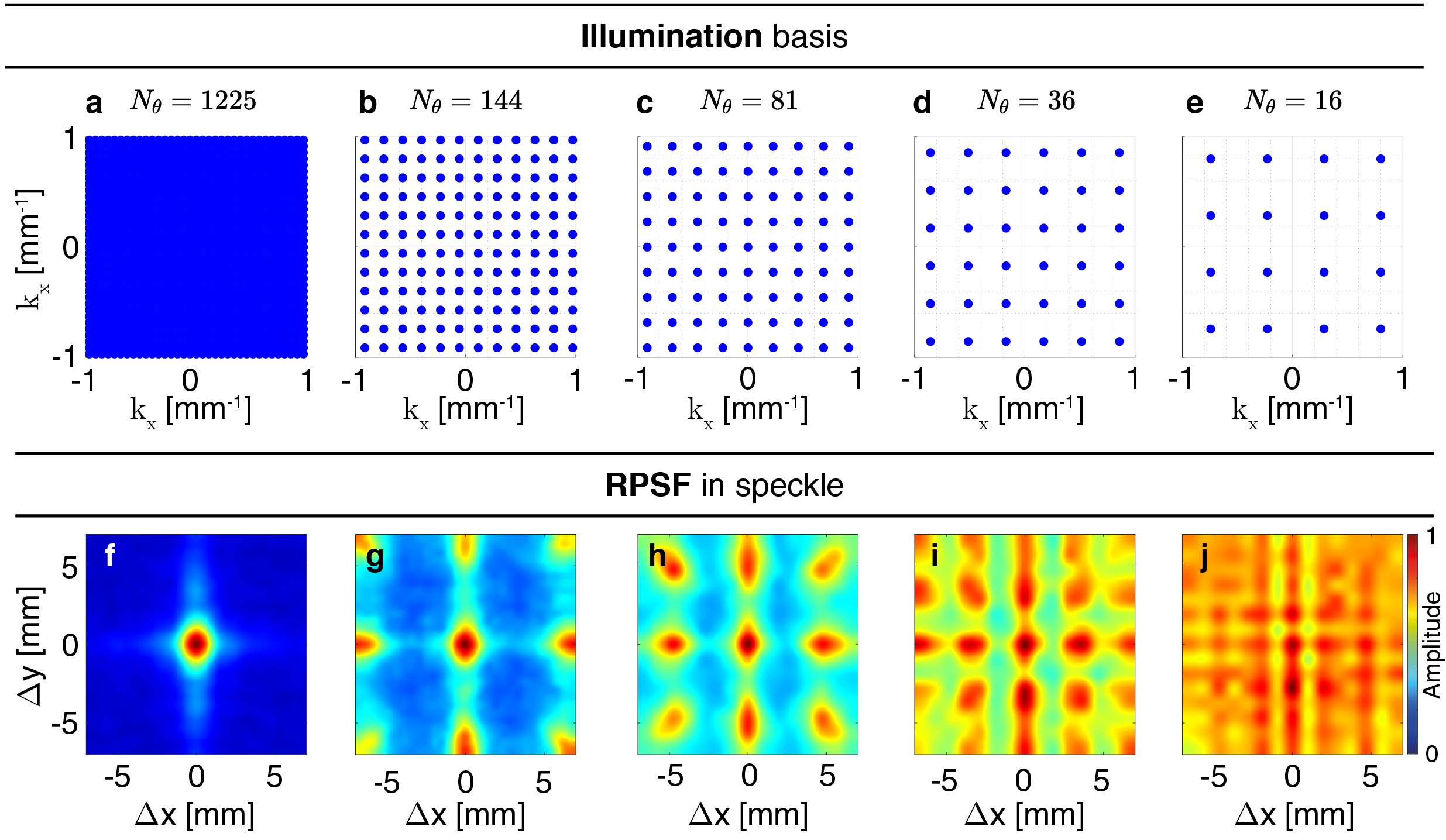}
 \caption{\textbf{Illumination sequence.} (\textbf{a}-\textbf{e}) Representation of different plane wave illumination sequence in the $\mathbf{k}-$space. (\textbf{f}-\textbf{j}) Aliasing effect exhibited by the RPSFs due to incompleteness of illumination sequence displayed in panels a-e, respectively. These RPSFs have been measured in a speckle area of a tissue-mimicking phantom.}
 \label{S7_Aliasing}
\end{figure}

Supplementary Figure~\ref{S7_Aliasing} shows the impact of the illumination sequence on UMI. If the input illumination basis is complete [Supplementary Figure~\ref{S7_Aliasing}a], the RSPF exhibits the expected diffraction-limited resolution [Supplementary Figure~\ref{S7_Aliasing}f]. The side lobes along the y-axis are due to the probe geometry made of four blocks of transducers separated by a distance of 0.5 mm ({three inactive rows of transducers along the y-axis}). 

When the number of illuminating plane waves is reduced [Supplementary Figures~\ref{S7_Aliasing}b-e], spatial aliasing occurs on corresponding RPSFs [Supplementary Figures~\ref{S7_Aliasing}g-j]. The maximal extension $\Delta \rho_{\mathrm{max}}$ of the RPSF has to be fixed to avoid the spatial aliasing induced by the incompleteness of the plane wave illumination basis; $\Delta \rho_{\mathrm{max}}$ is inversely proportional to the angular step $\delta \theta$ of the plane wave illumination basis: 
\begin{equation}
\label{aliasing}
\Delta \rho_{\mathrm{max}}\sim \lambda_{c}/(2\delta \theta)
\end{equation}
with $\lambda_c$ the central wavelength and $\delta \theta$ the angular pitch used for the illumination sequence. Thus, to avoid spatial aliasing, the coefficients $R(\bm{\rho}_\textrm{in},\bm{\rho}_\textrm{out},z)$ associated with a transverse distance $|\bm{\rho}_\textrm{out} - \bm{\rho}_\textrm{in}|$ larger than the superior bound $\Delta \rho_{\mathrm{max}}$ should be filtered via a confocal filter. 

Equation~\ref{aliasing} implies the necessity of recording a high-dimension $\mathbf{R}-$matrix for transcranial imaging, as aberrations are particularly important in that configuration (see Fig.~5 of the accompanying paper). The number of independent incident waves should scale as the number of resolution cells over which the RPSF spreads.

\section{Discriminate multiple scattering from electronic noise}
{We consider here the background of the focused reflection matrix for a given point $\rgp$:
\begin{equation}
    B(\drho,\rg_p)=\langle R_{\mathcal{M}}(\drho,\rgm)\mathcal{D}(\drho)\mathcal{W}(\rgm-\rg_p) \rangle_{\rgm}
\end{equation}
where $\mathcal{D}(\Delta \bm{\rho})$ is a de-scanned window function that eliminates the confocal peak and $\mathcal{W}$ is a spatial average window function around the targeted focal point $\rgp$.}

{The background can be decomposed as the sum of a fully symmetric matrix associated to multiple scattering (due to spatial reciprocity) and a fully random matrix associated to the electronic noise as follows:
\begin{equation}
    \underbrace{\mathbf{B}}_{\textrm{Background}}=\underbrace{\mathbf{M}}_{\textrm{Multiple scattering}}+\underbrace{\mathbf{N}}_{\textrm{Noise}}
\end{equation}
Projecting the $\mathbf{B}-$matrix onto its anti-symmetric subspace directly holds the anti-symmetric part of the electronic noise such that:
\begin{equation}
   \mathbf{B}^{(\textrm{A})}=\frac{\mathbf{B}-\mathbf{B}^\top}{2}=\mathbf{N}^{(\textrm{A})}
\end{equation}
Assuming equi-repartition of the electronic noise onto its symmetric and anti-symmetric subspace leads to:
\begin{align}
   \lVert \mathbf{B}^{(\textrm{A})} \rVert^2 =\lVert \mathbf{N}^{(\textrm{A})}\rVert^2=\frac{1}{2}\lVert \mathbf{N}\rVert^2
   \label{equirepart}
\end{align}
The norm of the background can be expressed as follows: 
\begin{equation}
    \lVert \mathbf{B} \rVert^2 = \lVert \mathbf{M} \rVert^2 + \lVert \mathbf{N} \rVert^2 +2 \underbrace{\langle \mathbf{M}|\mathbf{N}\rangle}_{\sim 0}
    \label{backgroundnoise}
\end{equation}
Assuming that the scalar product between the electronic noise and the multiple scattering is zero on average, the multiple scattering rate $\alpha_M$ can be derived by combining equations (\ref{equirepart}) \& (\ref{backgroundnoise}):
\begin{align}
    \alpha_M=\frac{\lVert \mathbf{M}\rVert^2}{\lVert \mathbf{B} \rVert^2}=&1-2\underbrace{\frac{\lVert \mathbf{B}^{(\textrm{A})} \rVert^2}{\lVert \mathbf{B} \rVert^2}}_{\beta}
    \label{multiprateanti}
\end{align}
with $\beta$ the anti-symmetric rate of the $\mathbf{B}-$matrix.}

\section{Notation and symbols}

\begin{table}[h!]
\center
\begin{tabular}{|c|c|}
  \hline
  \textbf{Symbol} & \textbf{Meaning} \\
  \hline
  \hline
  $\mathbf{R}$ & Reflection matrix \\
   % \hline
  %$\mathbf{R}_{\mathcal{M}}$ & Reflection matrix in a common midpoint basis \\
  \hline
      {$\mathbf{H}$} & {Point spread function matrix}\\
  \hline
       ${RPSF}$ & Reflection point spread function \\
     \hline
      {$\mathbf{G}$} & {Propagation matrix}\\
  \hline
  $\mathbf{D}$ & Distortion matrix\\
  \hline
  $\mathbf{C}$ & Correlation matrix\\
  \hline
  $\mathbf{\delta C}$& Perturbation term of $\mathbf{C}$\\
    \hline
  $\mathbf{T}$ and $\hat{\mathbf{T}}$  & Transmission matrix and its estimator \\
  \hline
  {{$| \delta {{T}}|^2$}}  & {Bias intensity of $\mathbf{{T}}-$matrix estimator} \\
  \hline
  {$P_{\textrm{in}/\textrm{out}}$} & {Scalar product between $\mathbf{\hat{T}}_{\textrm{in}}$ and $\mathbf{\hat{T}}_{\textrm{out}}$} \\
  \hline
   $\mathbf{i}$ & Illumination basis \\
  \hline
  $\mathbf{c}$ & Correction basis  \\
  \hline
 { $\mathbf{u}$} & {Transducer basis}  \\
  \hline
  {$\mathbf{k}$} & {Fourier basis } \\
    \hline
  {$\boldsymbol{\theta}$} & {Plane wave basis } \\
  \hline
 $l_c$ & Confocal filter size\\
  \hline
   ITR & Iterative Time Reversal\\
   \hline
   IPR & Iterative Phase Reversal \\
   \hline
   {$\mathbf{W}^{(n)}$} & {Wave-front of the ITR process at iteration $n$}\\
   \hline
  $\rg_m$ & Common midpoint\\
     \hline
  {$\rg_p$} & {Central point of a patch}\\
    \hline
  $\bm{\Delta \rho}=\rhog_{\textrm{out}}-\rhog_{\textrm{in}}$& Distance input/output focusing points \\
      \hline
  $\mathcal{D}(\drho)$& De-scanned window function\\
  \hline
\end{tabular}
\caption{\textbf{List of symbols for matrix imaging.}}
\label{NotationsUMI}
\end{table}

{\renewcommand{\arraystretch}{1.1}
\begin{table}[htb]
\centering
\begin{tabular}{|c|c|c|}
  \hline
 \textbf{Basis} & \multicolumn{1}{c|}{\textbf{Symbol}} & \textbf{Adapted for}\\
  \hline
  \hline
  {Acquisition basis} & {$\R_{\mathbf{iu}}(t)=[R(\mathbf{i}_{\textrm{in}},\mathbf{u}_{\textrm{out}},t)]$} &  Data recording\\
  \hline
  \hline
        \multirow{1}{*}{{{Focused} basis}}
    & \multirow{1}{*}{$\R_{\rhog\rhog}(z)=[R(\rhoin,\rhoout,z)]$}
        &    Focusing quality and multiple\\ 
       %\cline{1-2}
        \multirow{1}{*}{{{Common midpoint}}} & {$\R_{\mathcal{M}}(z)=[R(\drho,\rhog_\textrm{m},z)]$} &    scattering quantification \cite{lambert_ultrasound_2022}\\
    \hline
    \hline
    & $\R_{\mathbf{c}\rg}=[R(\mathbf{c}_{\textrm{in}},\rout)]$& \multirow{8}{*}{{{Local} aberration compensation  \cite{lambert_ultrasound_2021}}}\\
      {{Dual} basis}  & $\D_{\mathbf{c}\rg}=[D(\mathbf{c}_{\textrm{in}},\rout)]$ &  \\
      (input) &  $\C_{\textrm{in}}=[C(\mathbf{c}_{\textrm{in}},\mathbf{c}_{\textrm{in}}^\prime)]$ & \\
       &  $\mathbf{\hat{T}}_\textrm{in}=[\hat{T}(\mathbf{c}_{\textrm{in}},\rgp)]$  & \\
       
           \cline{1-2}
           
    & $\R_{\rg\mathbf{c}}=[R(\rin,\mathbf{c}_{\textrm{out}})]$ & \\
      {{Dual} basis}  & $\D_{\rg\mathbf{c}}=[D(\rin,\mathbf{c}_{\textrm{out}})]$ &  \\
      (output) &  $\C_{\textrm{out}}=[C(\mathbf{c}_{\textrm{out}},\mathbf{c}_{\textrm{out}}^\prime)]$  & \\
       &  $\mathbf{\hat{T}}_\textrm{out}=[\hat{T}(\rgp,\mathbf{c}_{\textrm{out}})]$ & \\
  \hline
\end{tabular}
\caption{{\textbf{Matrix notations.}}}
\label{RmatrixBasis}
\end{table}

\begin{table}[h!]
\center
\begin{tabular}{|c|c|}
  \hline
    \textbf{Symbol} & \textbf{Meaning} \\
    \hline
    \hline
  $\times$ & Matrix product \\
   \hline
   $\circ$ & Hadamard product\\
   \hline
   $\circledast$ & Convolution product \\
    \hline
   $\dagger$ & Transpose conjugate of a matrix \\
   \hline
      $\top$ & Matrix transpose \\
      \hline
   $\hat{}$ & Estimator of a physical quantity \\
   \hline
   SVD & Singular Value Decomposition \\
   \hline
    $\mathbf{U}^{(i)}$ & $i^{th}$ right singular vector of a matrix \\
       \hline
    $\mathbf{V}^{(i)}$ & $i^{th}$ left singular vector of a matrix \\
       \hline
    $\sigma_i$ & $i^{th}$ singular value of a matrix \\
       \hline
   $\langle \cdots \rangle$ & Ensemble average\\
  \hline
\end{tabular}
\caption{\textbf{Mathematical symbols.}}
\label{NotationsMath}
\end{table}

\begin{table}[h!]
\center
\begin{tabular}{|c|c|}
  \hline
  \textbf{Symbol} & \textbf{Meaning} \\
  \hline
  \hline
 $\mathcal{I}$ & {Image $\Leftrightarrow$ Estimation of the reflectivity}\\
  \hline
   $\rg=(x,y,z)$ & Focal point\\
  \hline
  $\bm{\rho}=(x,y)$ & Transverse coordinate\\
  \hline
 $\lambda_c$ & Wavelength at the central frequency\\
  \hline
$f_s$ & Sampling frequency\\
  \hline
  $f_c$ & Central frequency\\
  \hline
   $c_0$ & Speed-of-sound hypothesis\\
  \hline
  {$\mathbf{u}=(u_x,u_y,0)$} & Transducer position \\
  \hline
$\delta \rho_0$ & Transverse ideal resolution\\
  \hline
    $t$ & Time\\
      \hline
    {$\tau$} & {Time-of-flight}\\
      \hline
    {$\Delta \tau$} & {Time-delay}\\
  \hline
  $\gamma$ & Medium reflectivity\\
  \hline
  $\bm{\theta}=[\theta_x,\theta_y]$ & Plane wave\\
  \hline
    $\mathbf{k}=[k_x,k_y]$ & Fourier basis\\
   \hline
      $\beta$ & Anti-symmetric rate of a matrix\\
   \hline
   $\theta_{max}$ & Directivity of transducers\\
    \hline
   $\delta \theta$ & Plane wave sampling \\
    \hline
  $\mathbf{\Delta u}=(\Delta u_x,\Delta u_y)$ & Probe dimension\\
   \hline
    {$\mathcal{C}$} & {Coherence factor}\\
    \hline
    $\mathcal{A}_{(-3\textrm{dB})}$ & Area above 1/2 on RPSF amplitude\\
   \hline
   $\delta \rho_{(-3\textrm{dB})}$ & Experimental RPSF resolution\\
      \hline
   $\delta \rho_{0}$ & Diffraction-limited resolution\\
   \hline
    ${\mathcal{F}}$ & RPSF contrast\\
   \hline
    $\alpha_\textrm{S}$ & RPSF single scattering rate\\
   \hline
   $\alpha_\textrm{M}$ & RPSF multiple scattering rate\\
   \hline
    $\alpha_\textrm{N}$ & RPSF electronic noise rate\\
       \hline
    $\alpha_\textrm{B}$ & RPSF background rate\\
   \hline
   $\mathcal{W}$ & Spatial average window function \\
   \hline
    $N_{\mathcal{W}}$ & Number of resolution cells in $\mathcal{W}$\\
   \hline
  $\mathbf{w}=(w_{\rho},w_z)=(\{w_x,w_y\},w_z)$ & Dimension of $\mathcal{W}$\\
  \hline
  $A$ & Apodization term of synthetic aperture\\
  \hline
\end{tabular}
\caption{\textbf{List of general symbols.}}
\label{NotationsGeneral}
\end{table}

\end{document}